# A First-Order Non-Homogeneous Markov Model for the Response of Spiking Neurons Stimulated by Small Phase-Continuous Signals

**Authors:** J. Tapson[1], C. Jin[2], A. van Schaik[2], and R. Etienne-Cummings[3]

**Author's Addresses:**
1. Department of Electrical Engineering, University of Cape Town, Rondebosch 7701, South Africa.
2. School of Electrical and Information Engineering, University of Sydney, Darlington, NSW 2006, Australia.
3. Department of Electrical and Computer Engineering, The Johns Hopkins University, Baltimore, MD 21218, USA.

---

## Abstract

We present a first-order non-homogeneous Markov model for the interspike-interval density of a continuously stimulated spiking neuron. The model allows the conditional interspike-interval density and the stationary interspike-interval density to be expressed as products of two separate functions, one of which describes only the neuron characteristics, and the other of which describes only the signal characteristics. The approximation shows particularly clearly that signal autocorrelations and cross-correlations arise as natural features of the interspike-interval density, and are particularly clear for small signals and moderate noise. We show that this model simplifies the design of spiking neuron cross-correlation systems, and describe a four-neuron mutual inhibition network that generates a cross-correlation output for two input signals.

## 1. Introduction

In recent years there has been considerable effort applied to the challenge of modeling the response of integrate-and-fire neurons to periodic stimuli and noise (Tuckwell, 1988; Bulsara et al., 1994; Bair & Koch, 1996; Plesser & Tanaka, 1997; Plesser, 1999; Burkitt & Clark, 2000; Plesser & Geisel, 1999; Plesser & Gerstner, 2000; Svirskis & Rinzel, 2000; Amemori & Ishii, 2001; Gerstner & Kistler, 2002). To some extent this has been triggered by a broad interest in stochastic resonance (Bulsara et al., 1994; Plesser & Tanaka, 1997; Gammatoini et al., 1998), but more generally it has been motivated by the desire to have accurate models for a realistic range of neural behavior. When neurons are stimulated with periodic signals, there are a number of qualitatively different patterns of response. For large signals and low noise, phase-locking between the spiking output and the stimulus is very likely, particularly for subthreshold neurons; for superthreshold neurons, more complex effects may occur. Given the importance of phase-locking in some models of the sensory system, there have been many physiological and theoretical studies on this regime (Tuckwell, 1988; Bair & Koch, 1996). As the signal amplitude reduces and the noise increases, the neuron enters a regime where phase-locking is less likely, but there is still very strong synchronization between the signal features and the spikes. It is in this regime that stochastic resonance tends to be of interest, and has been extensively studied and modeled (Bulsara et al., 1994; Burkitt & Clark, 2000; Plesser & Tanaka, 1997; Gammatoini et al., 1998; Plesser & Geisel, 1999; Shimokawa et al., 1999).

At smaller amplitudes and higher noise, the neuron enters a regime where the spikes and the signal features are not coherent, but the distribution of the spikes is still affected by the signal. The model presented in this paper is most relevant in this regime. A number of prior models describe this type of behavior well (Plesser, 1999; Amemori & Ishii, 2001; Svirskis & Rinzel, 2000) but these models generally have a high burden of complexity in both theory and computation.

In the field of control systems, particular value is placed on models which allow a mathematical separation of system parameters and signal parameters, so that the response of a system to a signal can be characterized by a transfer function in which the signal is multiplied by or convolved with the system function. In modeling neurons, an analogy would be characterisation of the stimulated neuron response in terms of a product or convolution of the stationary un-stimulated neural response, and the signal.

With well-defined electrochemical relations for a neuron (such as the Hodgkin-Huxley equations), we can model its behavior to arbitrary degrees of accuracy using Monte-Carlo methods; and given the ever-increasing power of computer hardware, this becomes a progressively more attractive option. Under the circumstances, it is not unreasonable to ask why we need increasingly complex analytical models such as those referred to above, particularly as they inevitably involve a trade-off of accuracy for computational brevity. An analytical model is essentially a theoretical description of a neuron which may include internal variables and input-response relationships; for noisy neurons, it is likely to be inherently statistical in nature. The purposes of such a model include:

- Providing insight and a shorthand language for the dynamics of the neuron's behavior.
- Permitting rapid calculation of ensemble and time-averaged response statistics, without resorting to Monte-Carlo methods.
- Allowing the development of an intuitive understanding of neural behavior, from which educated and insightful predictions can be made.

In this paper we propose a first-order homogenous Markov model for the response of an integrate-and-fire neuron to a small signal in the phase-continuous regime, which allows separation of neuron and signal characteristics. By phase-continuous, we mean that the phase of the input signal is not reset after each spike; the signal itself may be discontinuous in the sense of not having a smooth or well-defined derivative (for example, it may be a train of spikes from an upstream neuron). These are sometimes referred to as *continuous* or *exogenous* or *unconstrained* signals, depending on which subtleties of the situation the authors wish to emphasize. While it is generally accepted that real-world sensory neurons are subject to signals of this type, in many studies which have considered the signal-processing properties of spiking neuron systems, each reset-to-firing event is treated as a separate and independent trial, in order to reduce the analytical complexity (Tuckwell, 1988; Rieke et al., 1996; Gerstner & Kistler, 2002). The use of peri-/post-stimulus time histograms (PSTHs) also encourages the analysis of data with respect to fixed stimulus phases.

In the research which follows, we consider successive trials to be non-independent, and begin each new integrate-to-firing trial at the phase at which the previous trial ended (fired). The phase is therefore a system variable that is "remembered" from one trial to the next (because it is a function of the absolute time $t$). We are aware of a small number of neuron models which explicitly accommodate phase continuity (Plesser, 1999; Burkitt & Clark, 2000; Plesser & Geisel, 1999; Shimokawa et al., 1999b; Amemori & Ishii, 2001). The model which we propose retains phase continuity, but differs from these prior models in being simpler; allowing a separation of neural and signal variables; and encouraging a useful intuitive understanding of neurons operating in this regime.

In developing the model, we will start by offering a brief mathematical justification for the approach, and show the range of parameters over which its accuracy has been established by comparison with Monte-Carlo simulations (i.e. the regime within which it is an accurate model). We demonstrate its application with a variety of neuron and signal types, and conclude by illustrating its usefulness with the analysis of two simple neural circuits which implement correlation functions. In particular, we describe a neural circuit which arises from this model, which acts to cross-correlate two input signals over a wide range of inter-signal time delays.

It should be clear that we do not offer this model as an improvement in accuracy on prior models, but rather we suggest that it is more useful. Utility is more difficult to demonstrate than accuracy. We have tried to demonstrate this model's utility in two ways: firstly, by showing that it allows prediction of the effect of a small signal on a neuron, even when the underlying model for the neuron is unknown or analytically intractable; and secondly, by using the new information provided by the model to design from scratch a physiologically plausible neuron circuit with specific signal-processing capabilities.

In the work that follows we make several assumptions that reflect common simplifications for modeling purposes. We assume that the spike times of an unstimulated neuron describe a wide sense stationary time series and a stationary interspike-interval distribution. When we apply a stimulus, we represent it as a signal which is continuous in time and limited in amplitude and bandwidth, and consider this to represent the net effect of a large number of synaptically-coupled inhibitory and excitatory input spikes. We model noise as being Gaussian-distributed noise added directly to the membrane potential, regardless of its nominal physiological source (which may be diffusion, ion channel action, transmitter packet release, or any other stochastic process which may be continuous or quantal in nature). Although it is not important for the development of the model, we consider that the output of a single neuron over a long period of time is of interest, in that it throws light on the output of a population of neurons over a short period of time.

## 2. Modulation of the Probability of Firing Due to a Periodic Stimulus

We want to answer the following question: what effect does a continuous, periodic input stimulus have on the distribution of firing intervals for a spiking neuron?

Consider a single-compartment neuron whose membrane potential $v$, in the absence of stimulation, is a stationary time signal represented by the integral of some function $f(v,t,\sigma)$,

$$v(\tau) = \int_0^\tau f(v,t,\sigma)dt \tag{1}$$

where $\tau$ is the interval since the last firing event, at $t = 0$ in this case, and $\sigma$ is a noise parameter (for example, the standard deviation of some additive Gaussian white noise). The function $f$ may include refractory periods, leakages, drifts and various nonlinearities – most single-compartment models may be described in this way.

The firing pattern of the unstimulated neuron is described by a stationary distribution or density of firing intervals given by $\rho(\tau)$. Without further knowledge of $f(v,\tau,\sigma)$, we cannot make any statement about the shape of $\rho(\tau)$. We can to some extent predict the effect on the shape of $\rho(\tau)$ of an added stimulus $g(t)$. Consider $g(t)$ to be periodic with zero mean, and period $T$; to be bandwidth-limited; and to have an amplitude which is small in comparison to the firing threshold $\theta$, so that $|g(t)| \ll \theta$ for all $t$. The membrane potential becomes:

$$v(\tau) = \int_{t_0}^{t_0+\tau} f(v,t,\sigma) + kg(t)dt \tag{2}$$

where $g(t)$ is a current stimulus and $k$ represents the capacitance of the membrane; we may set $k = 1$ for clarity from here on.

In (2) above, we have to treat time explicitly in the potential, and the firing events depend on the time of the last firing event ($t_0$) as well as the interval since that event ($\tau$). Each interval is linked in time by the continuity of the signal $g(t)$. Consider now the effect of $g(t)$ on $\rho(\tau)$. It will modulate the slope of $v(t)$ and the steeper the slope, the greater the chance that the neuron will fire. As the potential approaches the threshold $\theta$ the rate of change of $v(t)$ is

$$\frac{dv(\tau)}{dt} = \frac{d}{d\tau}\int_{t_0}^{t_0+\tau} f(v,t,\sigma) + g(t)dt \tag{3}$$

$$\left.\frac{dv(\tau)}{dt}\right|_{v\to\theta^-} = f(v,t_0+\tau,\sigma) + g(t_0+\tau)\,. \tag{4}$$

The stimulus is an additive modulation on the slope of the unstimulated neuron. We cannot express this as an effect on the interspike-interval density without taking into account the explicit dependence on time, so the new $\rho(\tau)$ must be expressed as a density which is conditional on $t_0$ as well as $\tau$; it is now $\rho(\tau|t_0)$. From (3) and (4) it is plausible to approximate the time-dependent interspike-interval distribution as:

$$\rho_{AM}\left(\tau|t_0\right) = \rho\left(\tau\right)\left(1 + wg\left(t_0+\tau\right)\right) \tag{5}$$

where $w$ is a constant weight accounting for the relative amplitude of $g(t)$ and $\theta$. This approximation indicates that the signal has an additive amplitude-modulating effect on the original $\rho(\tau)$. The first-order, non-homogenous Markov nature of the process is clearly embodied in this approximation. We will elaborate more on these issues later, but at present, we consider the further development of (5). We use the subscript *AM* to indicate the use of the amplitude-modulated interspike-interval density approximation for the time-dependent interspike-interval distribution. Equation (5) is somewhat ad-hoc, but in the absence of a more explicit specification of the neuron function $f(v,t,\sigma)$ than (4), it is as precise as is possible.

Note that the *ansatz* (5) is dependent on the period of $g(t)$ being relatively short compared to the typical length of $\tau$ (hereafter $\langle\tau\rangle$), so that $\rho(\tau)$ is effectively constant across a period $T$ of $g(t)$; in other words $T \ll \langle\tau\rangle$. It is also of questionable causality, although the conventional definition of $\rho(\tau)$ requires the use of time bins $\Delta\tau$ in which causality is similarly complicated (an event happening late in the time bin affects the density amplitude of the whole bin). The increase in error as $T \rightarrow \langle\tau\rangle$ is shown in Section 2.

Significantly, we have made no assumption about the underlying neural structure and dynamics which give rise to $\rho(\tau)$; it could at this stage be any one of a wide variety of spiking neural models for which (4) is a reasonable description.

Some discussion of the Markov property is required. The continuous integrate-and-fire process is Markovian in $dv/dt$ and the continuous membrane voltage (without spiking) is the integral of $dv/dt$ and is not Markovian (see Cox & Miller, 1965, p. 227). The spiking process which resets the neuron produces a sequence of firing intervals $t_k$ - $t_{k-1}$ for the unstimulated neuron in which the probability of an interval does not depend on the starting time of that interval, and the density of intervals is time-invariant. The time-invariance or stationary property of the firing intervals arises because the noise process (if white noise) has an autocorrelation function which is a delta impulse at the origin, and the neuron is reset after each firing. When the neuron is stimulated by a time-dependent stimulus, the time invariance is broken, and the probability of a spike interval becomes dependent on the starting time of that interval (Plesser & Geisel, 1999; Shimokawa et al., 1999). The sequence of firing intervals of the stimulated neuron is therefore a first-order Markov process (because $t_k$ depends only on $t_{k-1}$ and not $t_{k-2}$, $t_{k-3}$…) and it is non-homogenous because the probability of the firing interval is dependent on time. Strictly speaking, the first-order property only holds for neurons which experience a complete reset after each spike, and would not apply to neurons such as the Hodgkin-Huxley type, for example.

The time-dependent or conditional interspike-interval (cISI) density $\rho(\tau|t_0)$ is not as useful as a stationary interspike-interval (ISI) distribution $\rho(\tau)$, but marginalizing with respect to $t_0$ in order to get a stationary distribution in $\tau$ is not trivial. A route forward does present itself, however. We observe that the probability of firing for a specific interval $\tau = t_{k+1} - t_k$ depends on the joint probability of starting at $t_k$, and firing at $t_{k+1}$. The probability of starting at $t_k$ is the probability of the previous interval firing at $t_k$, and so is modulated by the amplitude of $g(t_k)$. The probability of firing at $t_{k+1}$ is modulated by $g(t_{k+1})$; so $\rho(\tau)$ for the stimulated neuron must be some function of the values of $g(t)$ separated by intervals of $\tau = t_{k+1} - t_k$. Another way of expressing this is that probability of firing with a certain interval depends on the stimulus at both the start and the end of the interval. The stimulus in between these points does not affect the slope of the membrane voltage at this time; we are focusing on the limit as $v(t) \rightarrow \theta$.

We note that the conditional densities $\rho(t_{k+1}|t_k)$ can be multiplied together to give the probability of a sequence of firing times $t_k$, $t_{k+1}$, $t_{k+2}$… so that for any train of spikes at times $\mathbf{t} = [t_1, t_2,\ldots t_n]$ following a reference spike at $t_0$, the spike train density $Y_n(\mathbf{t})$ is given by:

$$\begin{aligned} Y_n(\mathbf{t}) &= \prod_{k=1}^{n} \rho_{AM}\left(t_k - t_{k-1}|t_{k-1}\right) \\ &= \prod_{k=1}^{n} \rho\left(t_k - t_{k-1}\right)\left(1 + wg\left(t_k\right)\right) \end{aligned} \tag{6}$$

There are many different possibilities opened up by calculating marginal densities on $Y_n(\mathbf{t})$. For example, we can calculate the probability that the first and second spikes after $t_0$, that is to say $t_1$ and $t_2$, are separated by $\tau$. We take the sequence of conditional probabilities for firing at $t_1$ and then $t_2 = t_1 + \tau$, and marginalize over all possible values of $t_1$:

$$
\begin{aligned}
q_{2,1}(\tau, t_0) &= \{\text{Prob. after time } t_0 \text{ that first and second spike times are separated by } \tau\} \\
&= \int_{t_0}^{\infty} \rho_{AM}(t_1 + \tau | t_1) \rho_{AM}(t_1 | t_0) dt_1 \qquad (7) \\
&= \int_{t_0}^{\infty} \rho(\tau)\rho(t_1 - t_0)(1 + wg(t_1 + \tau))(1 + wg(t_1))dt_1 \\
&= \int_{t_0}^{\infty} \rho(\tau)\rho(t_1 - t_0)dt_1 + w\int_{t_0}^{\infty} \rho(\tau)\rho(t_1 - t_0)g(t_1 + \tau)dt_1 \\
&\quad + w\int_{t_0}^{\infty} \rho(\tau)\rho(t_1 - t_0)g(t_1)dt_1 + w^2\int_{t_0}^{\infty} \rho(\tau)\rho(t_1 - t_0)g(t_1 + \tau)g(t_1)dt_1 \qquad (8)
\end{aligned}
$$

Noting that $\rho(\tau)$ is a constant for fixed $\tau$, and that by the common definition of a probability density $\int_0^{\infty} \rho(t)dt = 1$; and that $\int_t^{t+T} g(t) = 0$ because $g(t)$ is periodic with period $T$ and zero mean; and that with the separation of timescales referred to previously, $\rho(\tau)$ is approximately constant over any interval $T$, we can approximate as follows:

$$
\int_{t_0}^{\infty} \rho(t_1 - t_0)g(t_1)dt_1 \approx \sum_{n=0}^{\infty} \rho(nT) \int_{t_0+nT}^{t_0+(n+1)T} g(t_1)dt_1 = 0,
$$

and

$$
\int_{t_0}^{\infty} \rho(t_1 - t_0)g(t_1 + \tau)g(t_1)dt_1 \approx \sum_{n=0}^{\infty} \rho(nT) \int_{t_0+nT}^{t_0+(n+1)T} g(t_1 + \tau)g(t_1)dt_1,
$$

so that (8) reduces to:

$$
\begin{aligned}
q_{2,1}(\tau, t_0) &= \rho(\tau) + w^2\rho(\tau)\int_{t_0}^{\infty} g(t_1 + \tau)g(t_1)dt_1 \\
&= \rho(\tau)\left(1 + w^2 R_{gg}(\tau)\right)
\end{aligned} \qquad (9)
$$

where $R_{gg}(\tau)$ is the autocorrelation function of $g(t)$, provided that the separation of timescales of $\rho(\tau)$ and $g(t)$ is sufficient to allow the use of these approximations. The interval of the integration can be reduced to the length of one period of $g(t)$ and normalized in the usual way.

In (9), $q_{2,1}(\tau,t_0)$ represents the distribution of the second spike interval after a spike at $t_0$, for all possible values of $t_1$; and as it has no explicit dependence on time apart from the signal $g(t)$, we can reasonably say that it represents the distribution of all first-order spike intervals after some first spike. Hence $q_{3,2}(\tau,t_1) = q_{2,1}(\tau,t_0)$ and so on (given the Markov property); we can therefore state $\rho_{AM}{}^{(s)}(\tau) = q_{2,1}(\tau,t_0)$. Note that this distribution is not applicable for the first spike interval $t_1 - t_0$ after some random start time $t_0$, and hence this distribution holds only for the second and subsequent (first-order) intervals in a continuously stimulated neuron.

### 2.1 Summary of the Model

This analysis gives us our basic amplitude-modulated interspike-interval density (AM-ISI) model for the time-dependent interspike-interval distribution. The model works as follows: for an unstimulated neuron, we assume there is a known stationary interspike-interval density, $\rho(\tau)$; when the neuron is stimulated by a periodic stimulus $g(t)$, we approximate the conditional and time-dependent interspike-interval density as shown previously (we repeat (5) here for clarity):

$$\rho_{AM}\left(\tau | t_0\right) = \rho\left(\tau\right)\left(1 + wg(\tau + t_0)\right). \tag{5}$$

We marginalize over $t_0$ as shown above and approximate the phase-continuous interspike-interval density with $q_{2,1}(\tau,t_0)$, to yield a stationary interspike-interval density:

$$\rho_{AM}{}^{(s)}\left(\tau\right) = \rho(\tau)\left(1 + w^2 R_{gg}\left(\tau\right)\right). \tag{10}$$

Some of the utility of the model may start to become apparent here. It suggests that, in the absence of any knowledge of a neuron structure except the unstimulated interspike-interval distribution, which is relatively easy to determine by observation, we can predict the effect of a continuous periodic stimulus on the interval after any given spike, and on the stationary interspike-interval distribution. We are not aware of any other neuron model which has this utility. In the following section, the range of parameters for which this approximation is accurate is examined.

Figure 1 gives a brief illustration of the model. The neuron used in Figure 1 is a simple integrate-and-fire neuron given by the function:

$$v(\tau) = \int_{t_0}^{t_0+\tau} m + \sigma\zeta(t) + kg(t)dt \tag{11}$$

where *m* is a constant drift, $\zeta(t)$ is additive white Gaussian noise with zero mean and unit variance, and $\sigma$ is the root-mean-square (rms) noise amplitude. Note that for *m*, $\zeta(t)$, and *g*(*t*) in current form, this equation has been normalized by setting *RC* = 1, where *R* and *C* are the membrane resistance and capacitance. For this simple case, there is a standard function for the unstimulated interspike-interval density given by, for example, Tuckwell (1988):

$$\rho(\tau) = \frac{\theta}{\sqrt{2\pi\sigma^2\tau^3}} \exp\left(-\frac{(\theta - m\tau)^2}{2\sigma^2\tau}\right). \tag{12}$$

We stimulate this neuron with a square wave signal, which we approximate mathematically as the sum of a Fourier series. This approximation has the benefit that we retain a finite, defined bandwidth and continuous derivative for the signal. The approximation appears to work for signals in which this is not the case, as shown in some later examples, but some of the mathematical analysis assumes these properties. The approximate square wave signal is given as:

$$kg(t) = A\sum_{n=0}^{n=9} \frac{\sin(2n+1)\omega_0 t}{2n+1} \tag{13}$$

so that

$$\rho_{AM}\left(\tau|t_0\right) = \left(\frac{\theta}{\sqrt{2\pi\sigma^2\tau^3}}\exp\left(\frac{(\theta - m\tau)^2}{2\sigma^2\tau}\right)\right)\left(1 + Aw\sum_{n=0}^{n=9} \frac{\sin(2n+1)\omega_0 t}{2n+1}\right) \tag{14}$$

and

$$\rho_{AM}{}^{(s)}(\tau) = \left(\frac{\theta}{\sqrt{2\pi\sigma^2\tau^3}}\exp\left(\frac{(\theta - m\tau)^2}{2\sigma^2\tau}\right)\right)\left(1 + A^2 w^2\sum_{n=0}^{n=9}\left(\frac{\sin(2n+1)\omega_0 t}{2n+1}\right)^2\right). \tag{15}$$

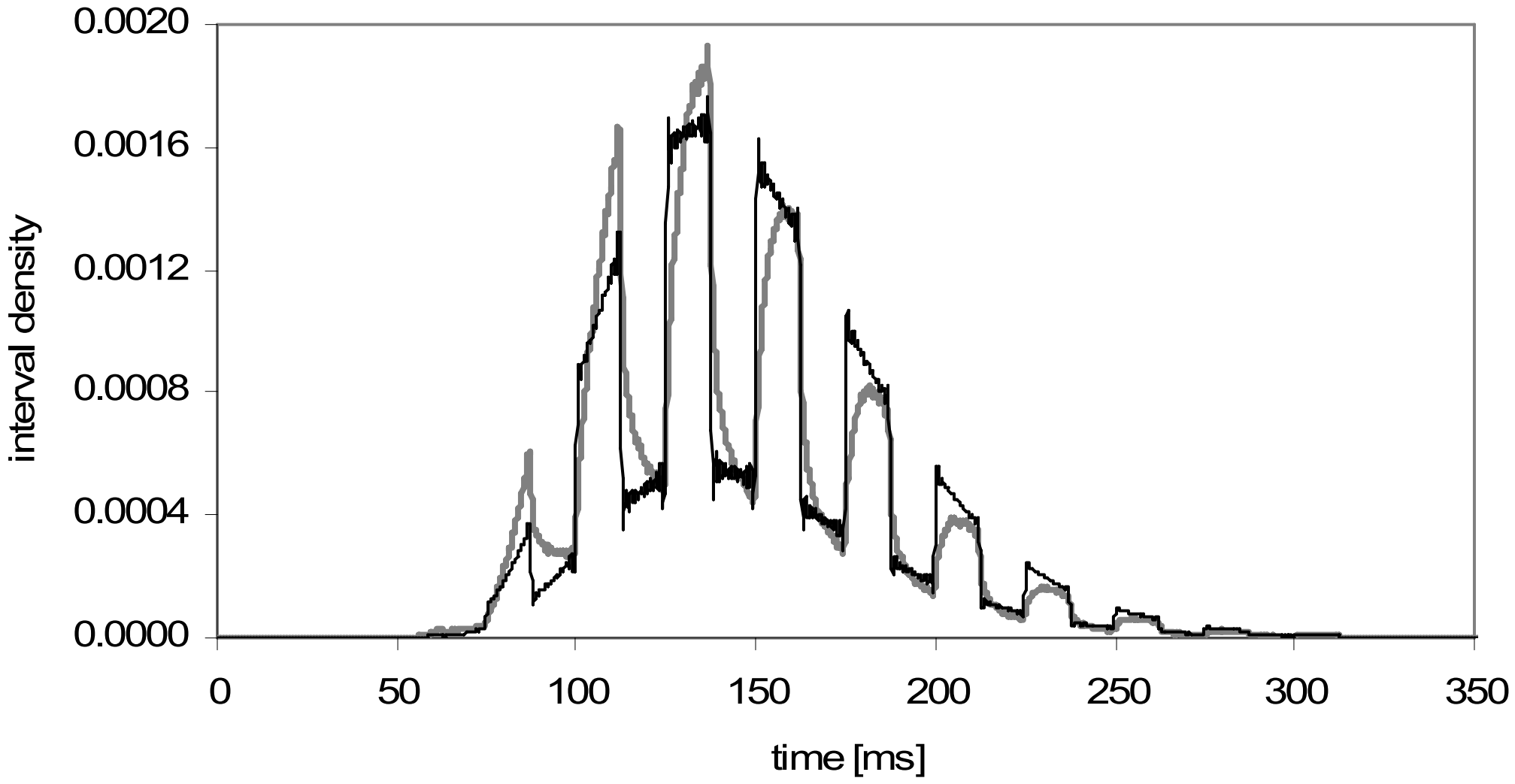

**Figure 1:** A comparison is shown of the conditional interspike-interval (cISI) densities obtained with the AM-ISI model (black) and a Monte-Carlo simulation (grey) for one value of the starting phase of the signal in (13), with $A$=100μV and $\omega_0$ = 80π rad/s. The cISI density of the AM-ISI model was calculated using (14). The neuron was a simple integrate-and-fire model with parameters $\sigma$=0.1 mV/√Hz, $\theta$=15 mV, $m$= 150 mV/s, $w$ = 6.25.

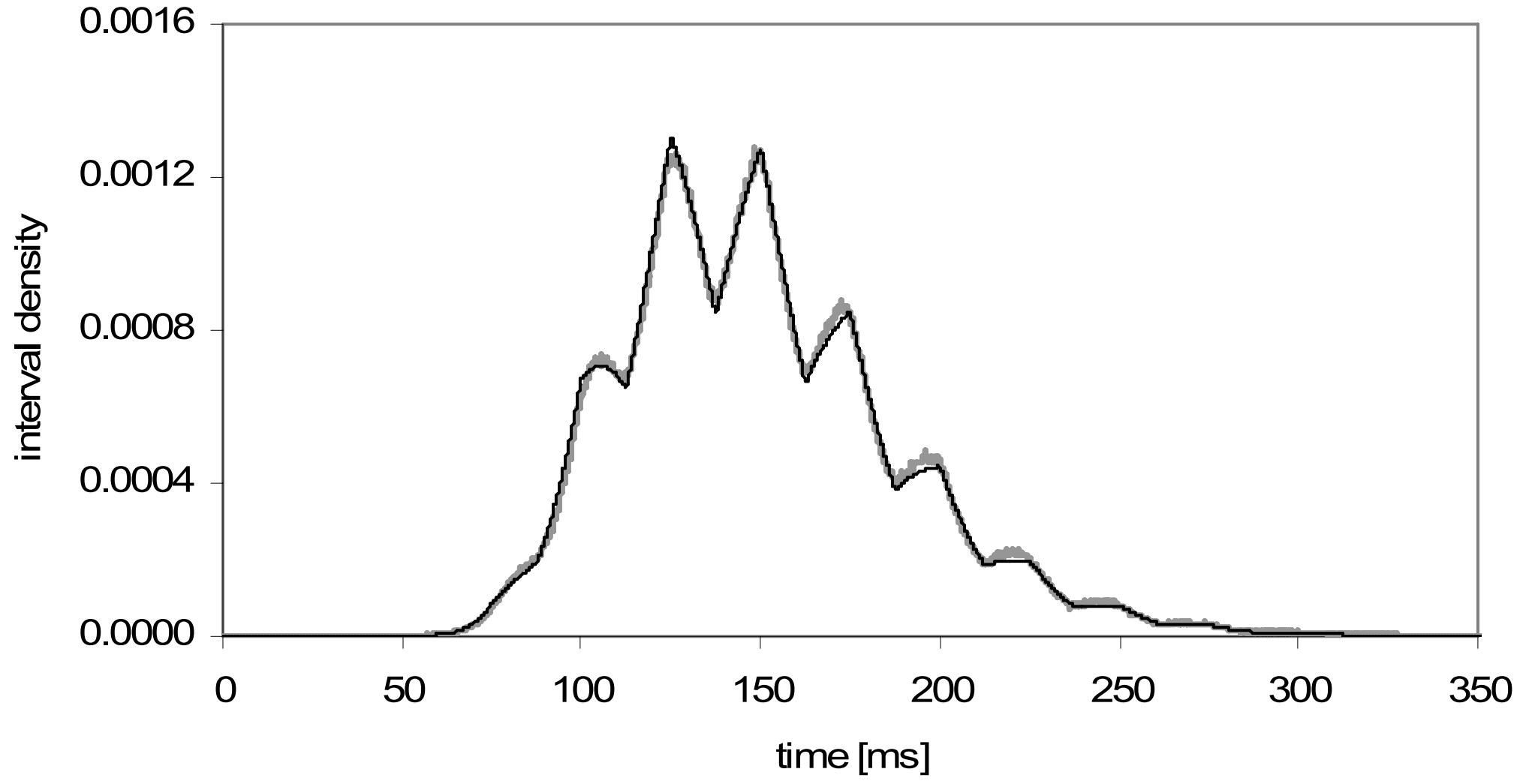

**Figure 2:** A comparison of the phase-continuous interspike-interval (ISI) density obtained with the AM-ISI model (black) and a Monte-Carlo simulation (grey) is shown for the signal in (13). The ISI for the AM-ISI model was calculated using (15). Note that the square wave signal has been converted into a triangular wave, which is the expected result of the autocorrelation operation. The neural parameters and signal were the same as in Figure 1.

In Figure 1, we show the non-phase-continuous case where the stimulus signal has been reset to some fixed phase value after each firing event. This is, therefore, a conditional interspike interval – it is the probability of a spike at τ given the signal state (phase) at time $t_0$. Figure 2 shows the phase continuous case, in which each interval started at the moment at which the previous interval ended (fired), apart from the trivial first interval, which started at $t = 0$; it is a stationary ISI. We can consider

Figure 1 to be the average of an ensemble of neurons, all of which started integrating at the same instant in response to the same signal; whereas Figure 2 represents the accumulation of thousands of firing events of a single continuously stimulated neuron. At risk of confusion, however, it must be pointed out that Figure 2 would not be significantly different if it were an ensemble average, provided that each neuron on the ensemble was firing continuously for some short period of time. The analysis in Eq.s 6-9 suggests that the ISI for the second first-order interspike interval (after a random start to initiate the first spike interval) is the same as the distribution for all first-order interspike intervals in continuous operation, and simulation results verify this. An ISI derived by accumulating a single interval from each of N neurons, is the same as one derived by accumulating N intervals from a single neuron, provided that for all cases the very first spike interval (which should have a random start time) is discarded.

We would like to draw the reader's attention to the clear autocorrelation effect displayed in Figure 2. This effect has been observed previously in measurements of all-order spike intervals from the mammalian auditory nerve (Cariani & Delgutte, 1996) and in interspike intervals in simulated neurons (Tapson, 1998); and has been implemented as a stochastic autocorrelation algorithm in electronic circuits (Tapson & Etienne-Cummings, 2007; Folowosele et al., 2007), but as far as we are aware this is the first analytical explanation for the effect; this prior work was based on the observed phenomenon, without any underlying mathematical basis being proposed. This model may also help to explain the effect of correlations in input spikes on the output statistics of integrate-and-fire neurons, as observed by several groups (see for example Salinas & Sejnowski, 2000; Kuhn et al., 2002).

### 2.2 Evaluation of the Accuracy of the AM-ISI Model

It is important to have a sense of how accurate the AM-ISI model is, in comparison to existing models and taking Monte-Carlo simulations as a baseline. There are a number of useful error metrics that we may use; for example, we may compare the measured (simulated) response with the model response, or with a measured response for the same neuron in the absence of a stimulating signal. In comparing spike probability densities, we have made use of the relative integrated mean square error:

$$E = \frac{\int_0^\infty d\tau[\rho_{measured}(\tau,t) - \rho_{model}(\tau,t)]^2}{\int_0^\infty d\tau[\rho_{measured}(\tau,t)]^2} \qquad (16)$$

where $\rho_{measured}(\tau,t)$ represents a measured or simulated interval density which may depend on time $t$ owing to the presence of a driving signal, and $\rho_{model}(\tau,t)$ is the corresponding model density. This error is intrinsically normalized by the denominator.

In Figure 3 we show the progression of accuracy of the model as the timescales of the neuron response and the periodic signal vary. As discussed earlier, the model is based on an assumption that the period of the signal is short relative to the expected value of $\tau$, so $T << \langle\tau\rangle$. Figure 3 shows the error $E$ for a simple integrate-and-fire neuron driven by periodic signals over the range $T/\langle\tau\rangle = 0.02$ to 3.00. It can be seen that the model is usefully accurate for $T/\langle\tau\rangle$ in the range (0.02, 1.50), after which the accuracy degrades. We note particularly that there is a minimum in error at approximately $T/\langle\tau\rangle = 1$. Shimokawa and colleagues have shown that the modulation of the ISIH of a leaky integrate-and-fire neuron by a phase continuous signal should be a maximum when the signal period is approximately equal to the ISIH peak (which we have denoted ‹τ›), although there is some variation from the 1:1 optimum of time scale matching as the neuron parameters change (Shimokawa et al., 1999b). The error minimum seen in Figure 3 is a function of the same effect, although here it is complicated by the use of a randomly-weighted harmonic series as a driving signal.

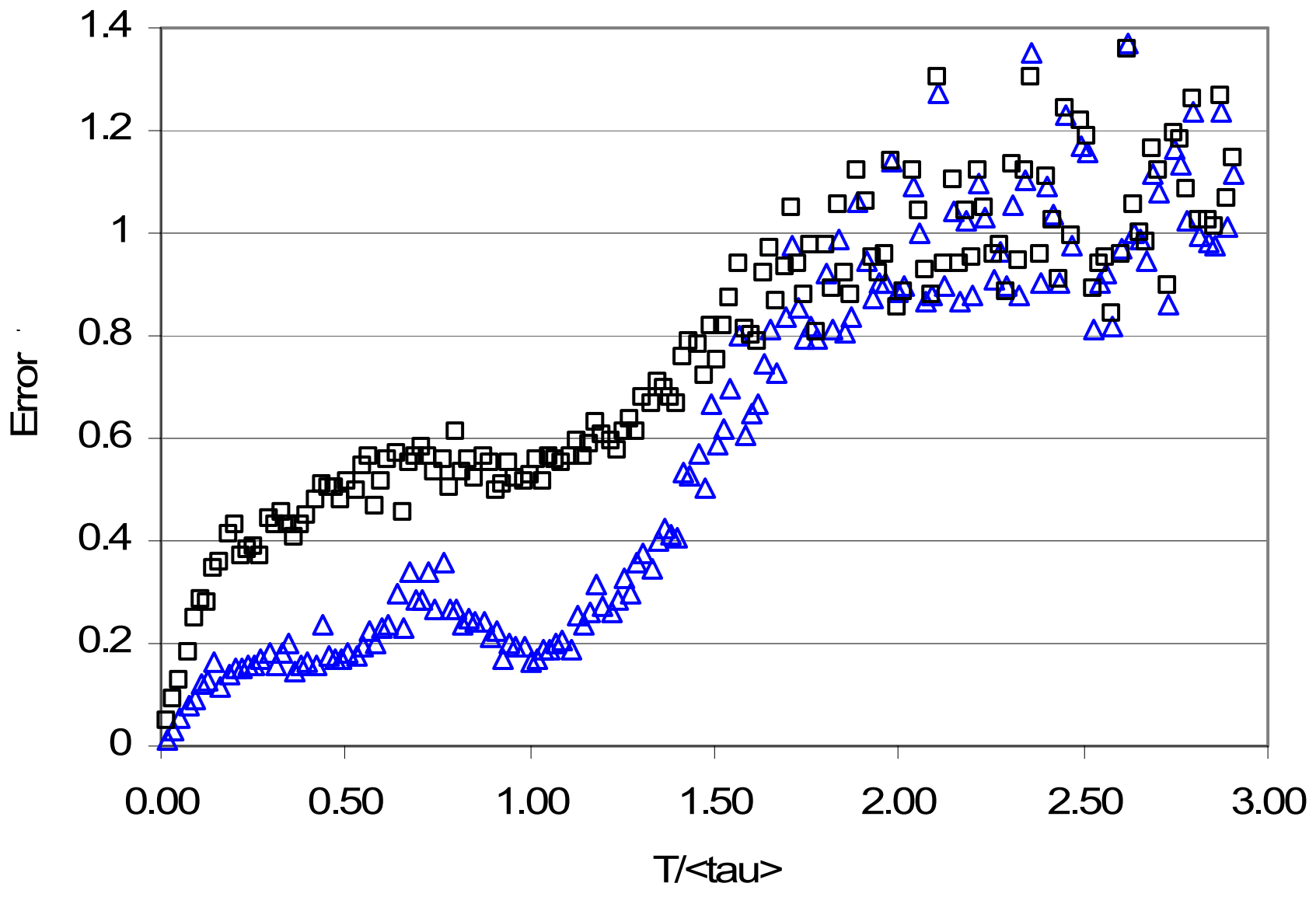

**Figure 3:** Error $E$ for a range of relative timescales $T/\langle\tau\rangle$ for a simple integrate-and-fire neuron driven by a periodic signal. The triangle symbols represent the error between the AM-ISI model (Eq. 10) and the measured (Monte-Carlo simulated) interval distributions. The square symbols represent the error between the standard undriven distribution (Eq. 12) and the measured interval distribution. The difference between these two curves essentially represents the improvement in accuracy gained by using the AM-ISI model. It can be seen that the AM-ISI model's utility diminishes above $T/\langle\tau\rangle = 1$. Each point represents the average of errors for twenty different test signals of the form $g(t) = \sum_{i=1}^{5} a_i \sin(\frac{i}{T}t + \phi_i)$ - a sum of the fundamental and first four harmonics of a series where the amplitudes $a_i$ and phases $\phi_i$ were randomly generated.

The neuron was a simple integrate-and-fire model with parameters $\sigma$=0.1 mV/√Hz, $\theta$=15 mV, $m$=100 mV/s and ‹τ› ≈ 137 ms.

We have evaluated the approximation for several neuron models, and in the case of the simple integrate-and-fire neuron, we have systematically explored a range of neuron and signal parameters. Signals were constructed by adding up to five Fourier components with phases and amplitudes chosen randomly within a defined range, and with frequencies chosen randomly from a set which was selected to give a combined signal with a period within a maximum range:

$$g(t) = \sum_{i=1}^{5} a_i \sin(\omega_i t + \phi_i) \tag{17}$$

$a_i$ a random number from a uniform distribution in [0,1]

$\omega_i$ randomly selected from the set

$\{\frac{1000}{nR}, n = 1,2,...10;$ R a random number uniformly distributed in [1, 1.5]}

$\phi_i$ randomly selected from a uniform distribution in [0, 2π].

For these tests, a single value of the optimization parameter $w$ in (9) was used, although it was adjusted by the noise variance (i.e. the constant was $w$=6/√$\sigma$). This value was arrived at by evaluating the error between the AM-ISI model and Monte-Carlo simulations for approximately 2000 combinations of input and neuron parameters. Optimization was carried out for the stationary and phase-continuous interspike interval condition (Eq. 10) as this is generally of more interest to practitioners (as well as generally being harder to model using other methods). The use of the constant was implemented throughout these tests.

Figure 4 shows the range of parameters within which $E$<0.01 for the AM-ISI model.

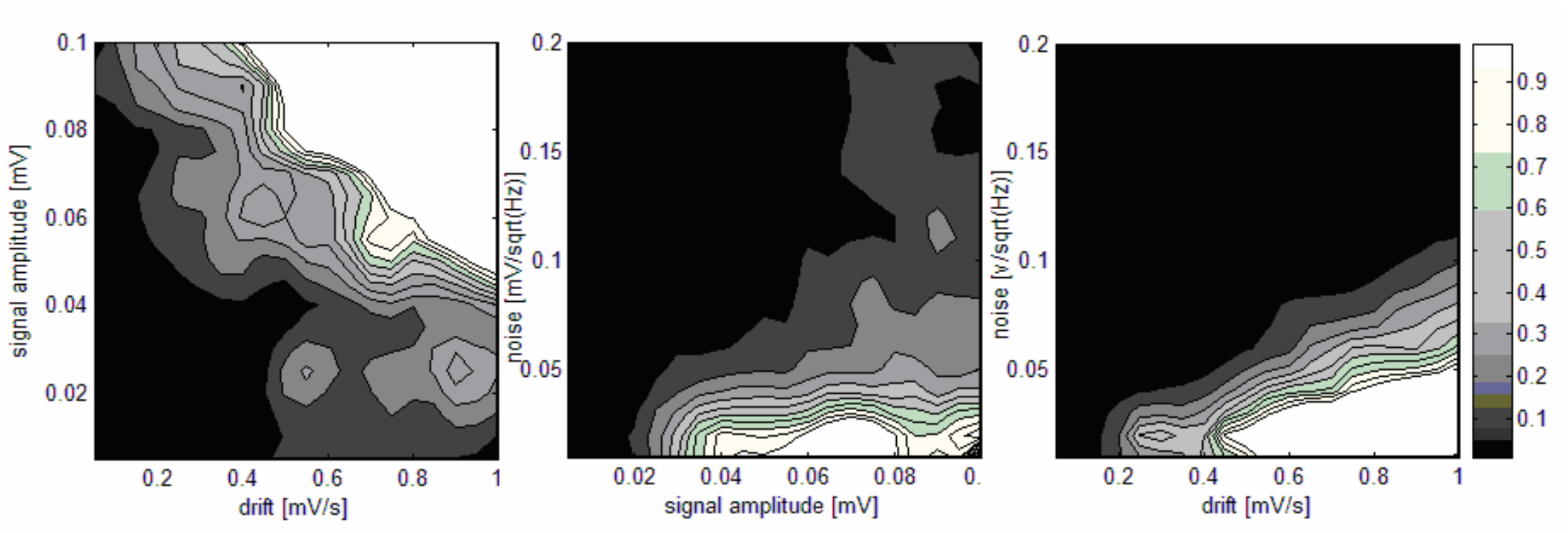

**Figure 4:** These graphs show contours of error, calculated using (16), for 2000 tests in which the integrate-and-fire neuron response to randomly generated periodic signals was modeled, with differing neural parameters. The neuron was a simple integrate-and-fire model with $\theta$ = 15 mV. The black areas have a relative integrated mean square error of $E < 0.01$; that is to say, less than 1% error. A single value of the weight parameter, adjusted for noise ($w=6/\sqrt{\sigma}$), was used in these tests. For the data in the left hand surface, $\sigma$=0.05 mV/√Hz; in the middle surface data, drift $m$ = 0.2 mV/s; and in the right hand surface data, the signal amplitude $A$ = 0.3 mV.

As seen in Figure 4, the error increases significantly when the signal amplitude is sufficiently large that $1+w^2g(t)/\sigma < 0$ for some $t$. This condition produces negative probability densities, unless we constrain $\rho_{AM}(\tau)$ = max[$\rho(\tau)$ $(1+w^2g(t)/\sigma)$ , 0]. Applying this constraint is not a complete solution as the normalization $\int_0^\infty \rho_{AM}(\tau)d\tau = 1$ is then required, and this in turn changes the optimal value of the weight $w$. Qualitatively, the curves produced by this procedure are a good match to the simulated ISIs (see Figure 7 for an example), but we have not evaluated the error as it would require optimization of $w$ for each test curve.

We conclude from the results in Figure 4, as well as from numerous examples similar to Figures 1-2, that there is a significant range of parameters for which this approximation is useful and accurate.

## 2.3. The Application of the Model to Different Neuron Types

In this section we demonstrate the model's utility by applying it to different types of spiking neuron models. The models to which we have successfully applied it include the simple integrate-and-fire neuron of Section 2.1 above (hereafter the I&F neuron); an Ornstein-Uhlenbeck (OU) neuron; and a Hodgkin-Huxley (HH) neuron. In addition to this, but not shown for reasons of brevity, we have also applied this model to the OU neuron with a refractory period; to a neuron with a quadratic nonlinear leakage term and refractory period; and to a neuron with an arbitrary nonlinear leakage consisting of piecewise linear leakages.

### 2.3.1 An Ornstein-Uhlenbeck Neuron

The Ornstein-Uhlenbeck neuron, which can also be described as a noisy, leaky integrate-and-fire neuron, is commonly regarded as a useful single-compartment model in its compromise of simplicity and accuracy. The membrane potential is typically described as follows:

$$\tau_m \frac{dv(t)}{dt} = -\left(v(t) - V_L\right) + i(t) + \sigma\zeta(t) \tag{18}$$

where $\tau_m = RC$, the membrane time constant, is the product of the membrane resistance $R$ and capacitance $C$; $V_L$ is the resting potential of the neuron; $i(t)$ is the stimulus signal and $\sigma$, $\zeta$ and $v$ are the rms noise amplitude, noise process, and membrane potential, as before. In addition, the model requires a defined threshold membrane voltage $\theta$ at which level the neuron will produce a spike; a reset potential $V_{reset}$, to which the membrane will be reset, which may or may not be equal to $V_L$; and an optional hyperpolarizing barrier, or reflective lower bound to $v(t)$: $v(t) \geq v_{hyp}$.

Figures 5 and 6 show the conditional interspike-interval (cISI) and phase-continuous interspike-intervals (ISI) respectively for an Ornstein-Uhlenbeck neuron implemented using commonplace values (Dayan & Abbott, 2001).

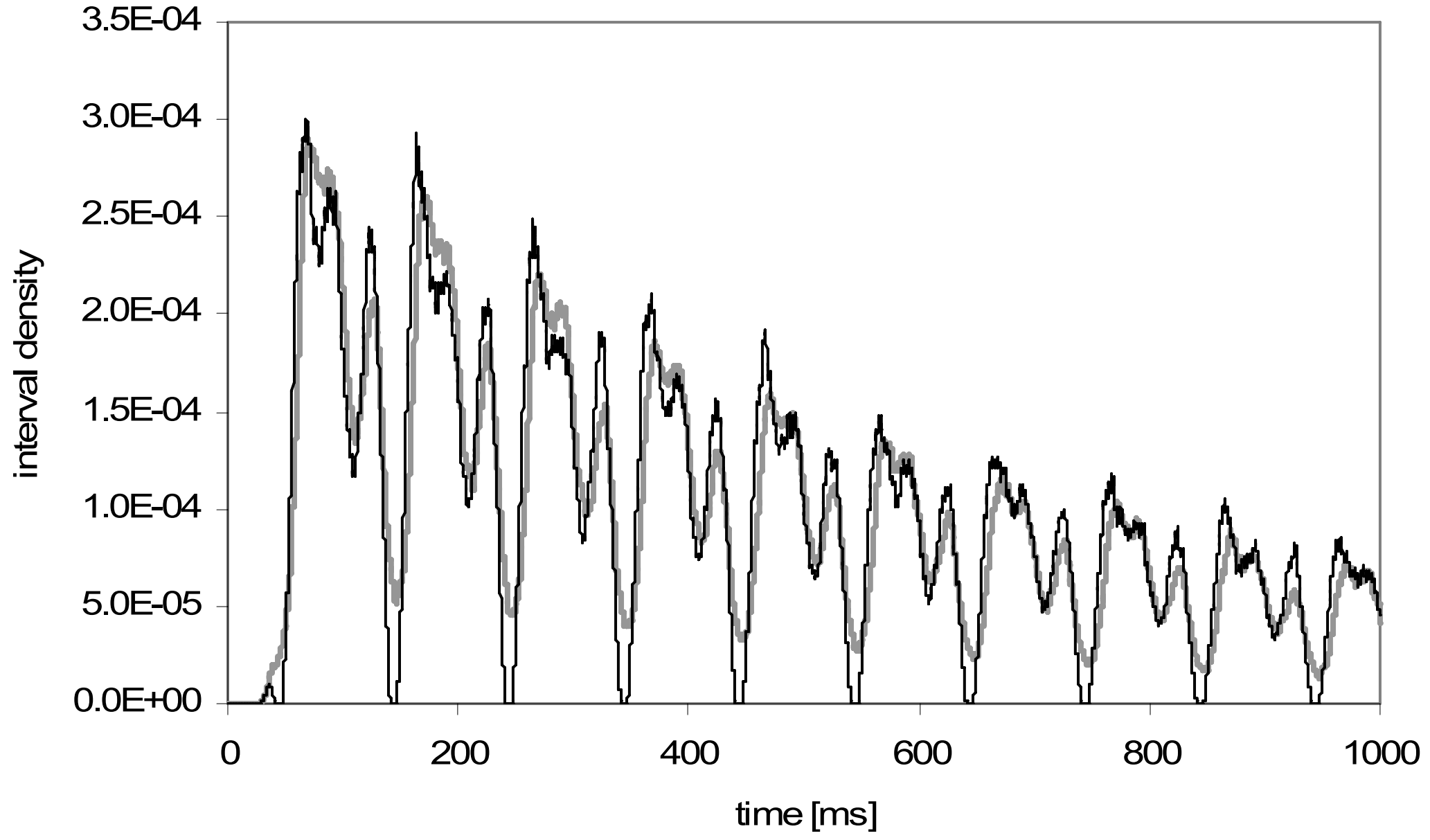

**Figure 5:** Conditional ISI for an OU neuron driven by an arbitrary periodic signal, with the AM-ISI model (black) and a Monte-Carlo simulation (grey). The signal was generated as described in Section 2.2, with $\omega_0 = 10\pi$ rad/s and with a constant drift component. Although there is some mismatch at low spike densities, the shape and periodicity of the model response is clearly appropriate. Neuron parameters were $\tau_m$ = 10ms, $\theta$ = -54mV, $V_{reset}$ = -80mV, $V_L$ = -70mV, $\sigma$=0.1 mV/√Hz, $w$ = 9.00.

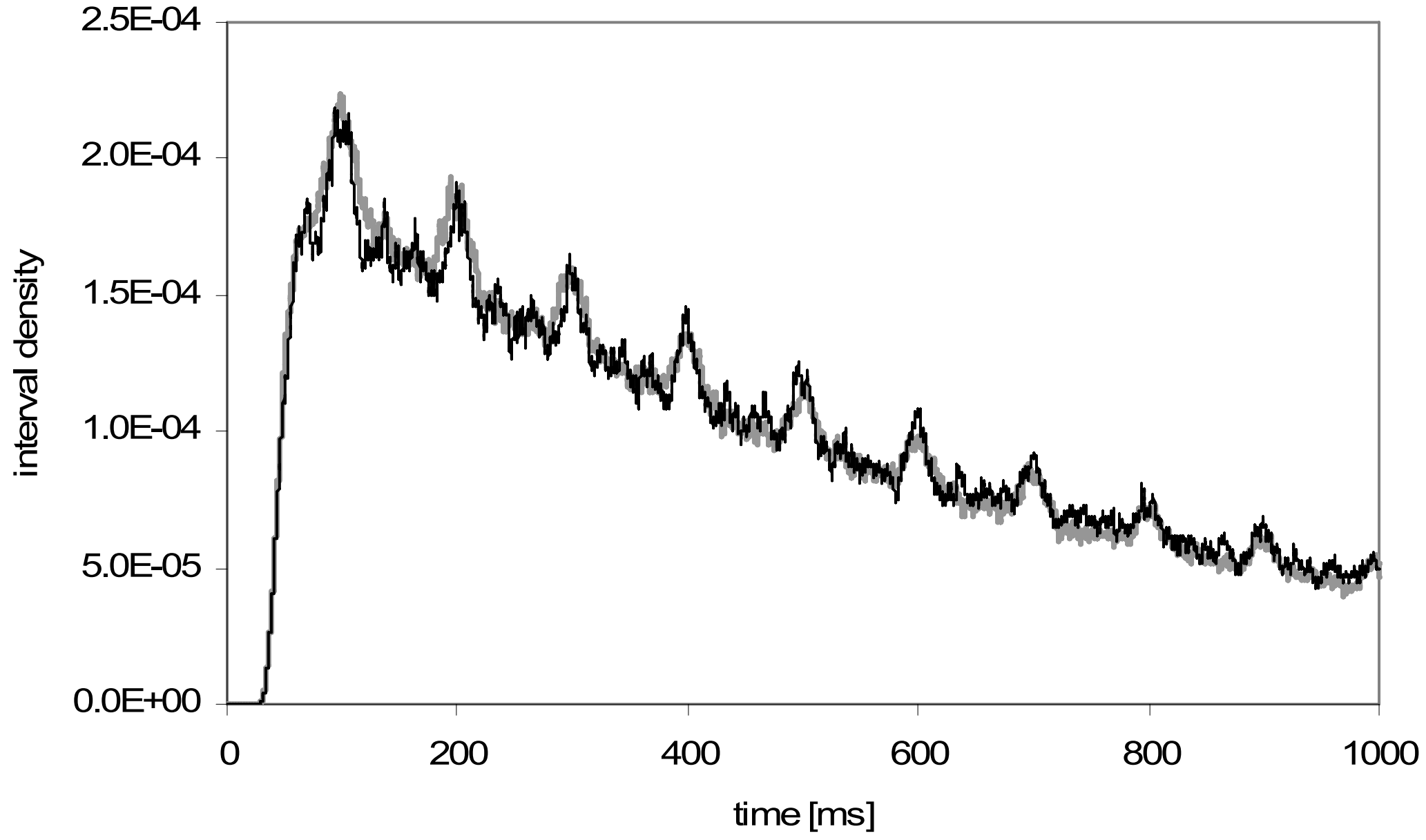

**Figure 6:** Phase-continuous ISI for an OU neuron with the AM-ISI model (black) and a Monte-Carlo simulation (grey), driven by an arbitrary periodic signal (the same neuron and signal as in Figure 5). The shape and position of features in the model and measured ISIs correspond well. The error (using (16)) between the curves was $E = 3.2 \times 10^{-3}$.

Figures 5 and 6 suggest that the approximation is a useful tool for predicting the shape of the ISI in the OU neuron, within an appropriate range of signal parameters. Note that in this example, the basic unstimulated ISI was obtained by measurement (using Monte-Carlo simulation) rather than by calculation, to demonstrate that *a priori* knowledge of neuron parameters is not necessary. That this approximation may still be used when the neuron model is unspecified, and is only accessible in terms of its unstimulated ISI, is a significant advantage which we elaborate on in the following section.

### 2.3.2 A Hodgkin-Huxley Neuron

The AM-ISI model may in principle be applied to any spiking neuron for which the response of the membrane potential near threshold can be modeled by Eq. 4. We have applied it to the standard Hodgkin-Huxley model (Hodgkin & Huxley, 1952; Dayan & Abbott, 2001) – the implementation is described in Appendix I. We show that the approximation is still useful for this model, despite the fact that the overall process is no longer strictly first-order Markovian (as the Hodgkin-Huxley neuron does not reset to an identical state after each firing event) and therefore the derivation in Section 2 above is less appropriate.

Figure 7 shows the cISI and Figure 8 the phase-continuous ISI for a Hodgkin-Huxley neuron driven by an arbitrary periodic signal.

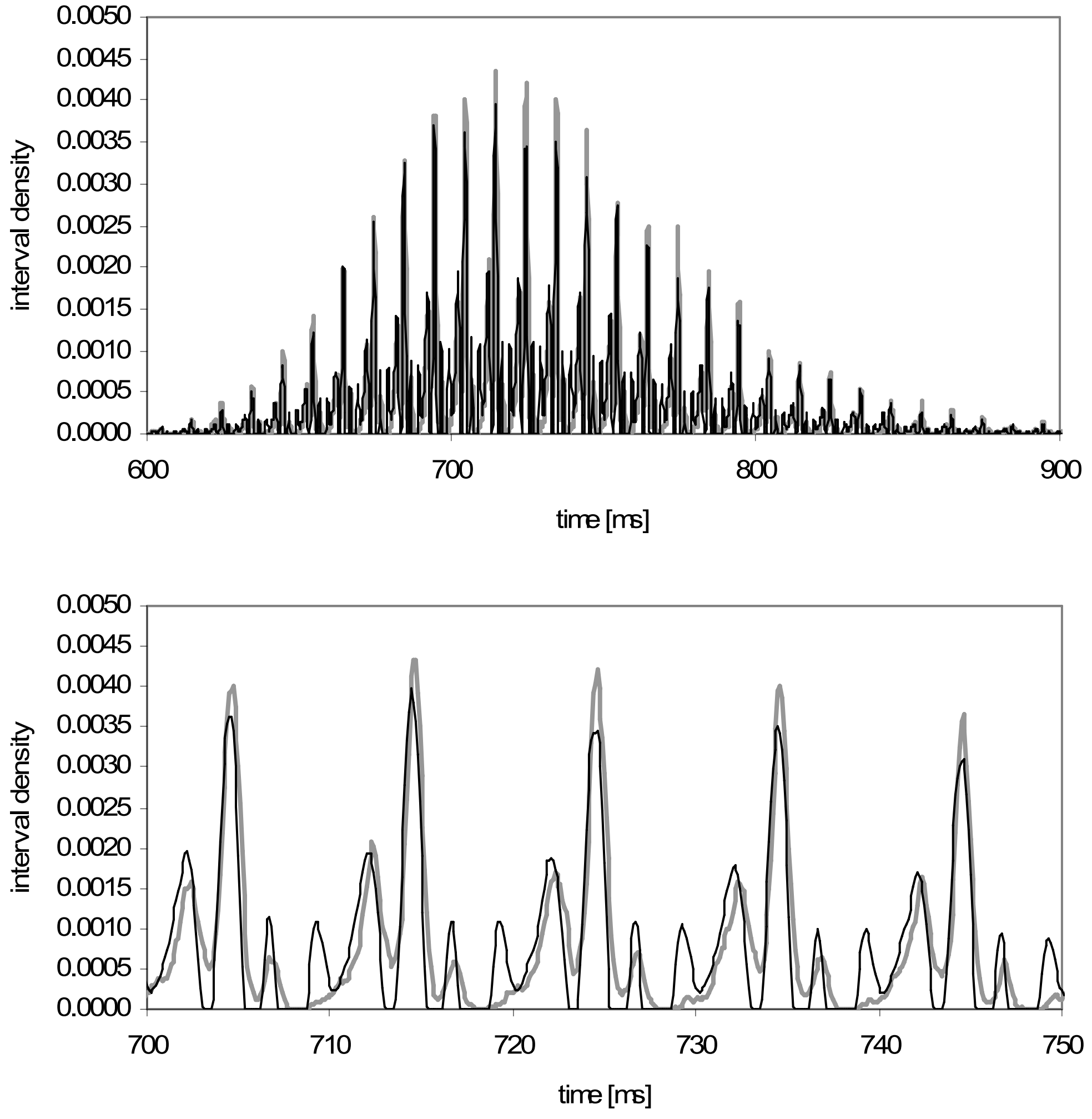

**Figure 7:** Conditional ISI for a Hodgkin-Huxley neuron driven by an arbitrary periodic signal, with the AM-ISI response model (black) and a Monte-Carlo simulation (grey). The upper graph shows the full non-zero spread of the distribution and the lower graph shows detail from the centre of the cISI. The signal was generated as described in Section 2.2, with $A = 2\times10^{-3}$ mA.cm$^{-2}$, $\omega_0 = 200\pi$ rad/s and with a constant drift component; $w$ = 13.0 for the model. The signal amplitude is at the upper limit for which the model is usefully accurate, as can be seen by the mismatch for intervals at which the densities are very low – the accuracy breaks down when there are regions in $\tau$ with no spikes; nonetheless, the salient features of the cISI are correctly modeled in respect of position and amplitude. The neuron parameters

were standard for the model (Dayan & Abott, 2001) and the implementation is described in Appendix I.

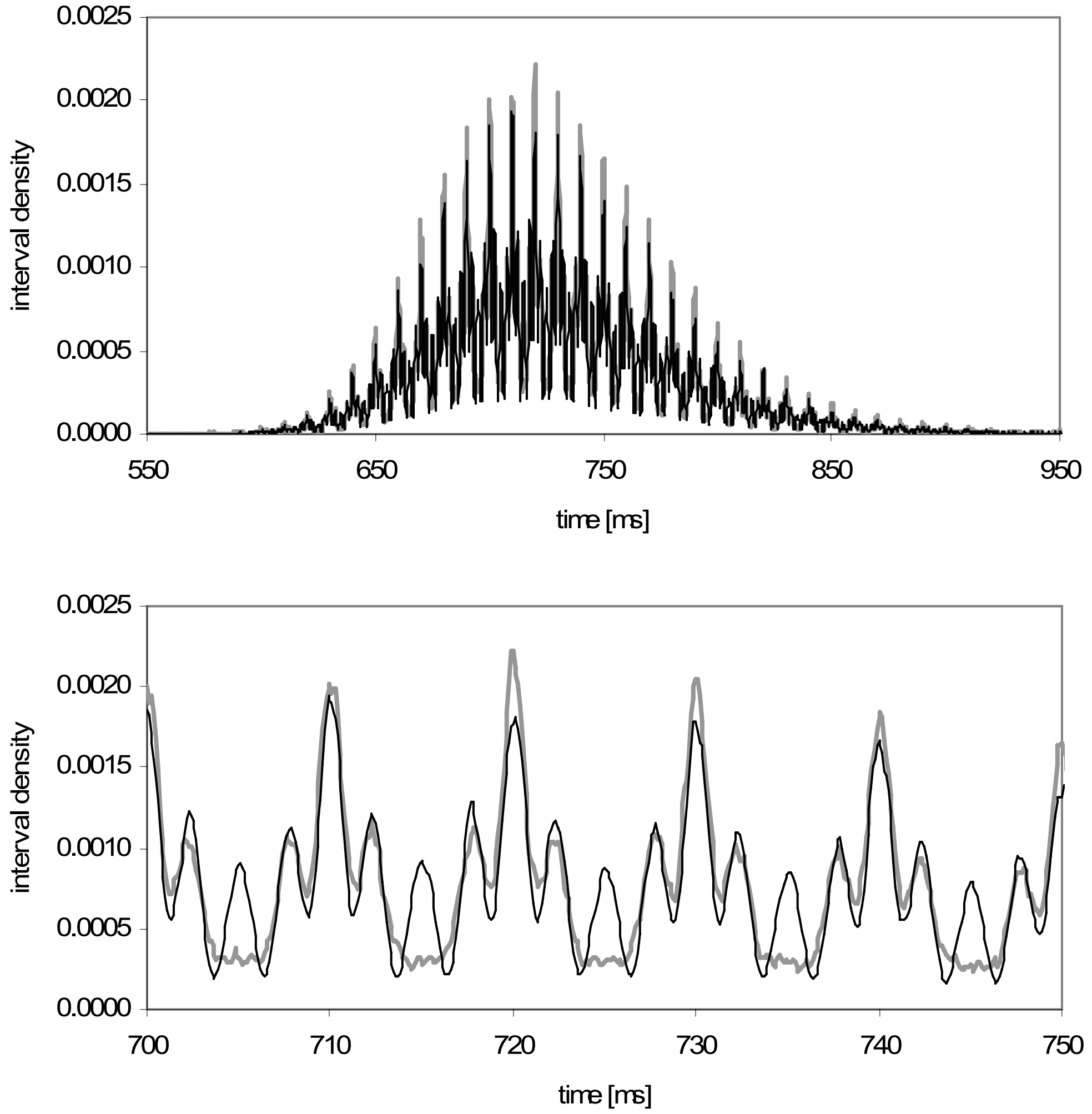

**Figure 8:** Phase-continuous ISI for a Hodgkin-Huxley neuron, as in Figure 7, driven by an arbitrary periodic signal, with the AM-ISI response model (black) and a Monte-Carlo simulation (grey). The upper graph shows the full non-zero spread of the distribution and the lower graph shows detail from the centre of the ISI. The signal was generated as described in Section 2.2, with $\omega_0 = 200\pi$ rad/s and with a constant drift component. Although there is some mismatch at low spike densities, the shape and periodicity of the model response is clearly appropriate. The absence of the lowest peak in the simulation results is, we believe, due to the approximation

becoming inaccurate when the negative signal slope exceeds the neuron drift in magnitude; larger than appropriate signals were used in these figures for visual clarity.

The success of the AM-ISI model in predicting the response of the Hodgkin-Huxley neuron demonstrates how the separation of the neuron and the stimulation into separate factors allows a useful prediction to be made, even when the underlying neural model is not analytically helpful. All that is required is the ISI of the unstimulated neuron, provided the signal meets the criteria in respect of timescales and amplitude.

### 2.4 Accuracy of the Autocorrelation Function

In a further example, we wish to draw attention to the accuracy of the autocorrelation product. In this case, we use a standard Ornstein-Uhlenbeck neuron (an integrate-and-fire neuron with ohmic leakage). As a stimulus, we use a pseudorandom noise (PRN) sequence. The use of PRN sequences requires some motivation as their use may not be well known to neuroscientists. PRN sequences are apparently random binary digital (bit) sequences which are constructed so that they have the appearance of noise, and they have sharp peaks at zero delay in the autocorrelation function; they are in that sense a binary approximation to delta-correlated white noise. However, it is possible to construct sets of PRN sequences which have very low mutual cross-correlation, i.e. they are orthogonal or nearly so. Hence, if we correlate one of these sequences with itself, we get a well-defined autocorrelation peak at zero delay, but if we cross-correlate it with a different member of the set, we get no significant cross-correlation peak at any delay. These sets of codes are used in many electronic systems to embed or encode information in a noise-like form, usually by multiplying the information with the PRN code, either in the time or frequency domain. This modulated signal is then transmitted for several epochs of the PRN code. The encoded information is recovered by cross-correlating the received signal with the full set of codes. Those codes which do not match to the original modulating code return low cross-correlations, whereas the code which does correspond to the modulating code produces a significant cross-correlation peak. This is the basis for the code-based multiplexing of signals in code-division multiple access (CDMA) telecommunications systems such as the GSM mobile telephone network. This is also the method by which time information is transmitted by the satellites of the Global Positioning System (GPS) –each satellite repeatedly broadcasts a unique code, and the receiver cross-correlates its input with local copies of all possible codes. The correlogram for each local code will contain a peak at a position corresponding to the time delay between the local code and the received code, and that time delay can be used to infer the time of flight of the signal from the satellite, and hence the distance to the satellite.

The particular PRN sequences we use are Gold codes, as used in the GPS system. Gold codes are a set of 36 PRN sequences, 1023 bits in length, which have been chosen to have extremely sharp autocorrelation peaks (one bit in width) and to be nearly orthogonal with each other, thereby making them very useful for CDMA communications systems (Sarwate & Pursley, 1980). The autocorrelation function of a PRN code should be relatively flat, with a single sharp peak at intervals corresponding to the 1023-bit code length.

Pseudorandom codes might seem at first sight to be peculiarly non-biological signals to use in a spiking neuron system. From our point of view they have some significant advantages. Their noise-like character is not dissimilar to the quasi-random spiking input that most neurons receive – there may be a weak argument that sine waves are more representative of the input to the auditory system, but most real sounds (and real auditory nerve signals) are quite irregular, and there are no useful analogies to the sine wave in vision input. More importantly, the sharp correlation peaks produced by PRN codes allow exact determination of the presence and accuracy of autocorrelation and cross-correlation terms; so we will introduce their use here and return to it in Section 3. In addition, the correlation peaks obtained when autocorrelating or cross-correlating PRN codes are independent of the signal envelope (which is flat) and the phase (which is random), hence emphasizing that what we obtain is a real mathematical correlation and not some strange artifact of the integrate-and-fire process. Another way of putting this is that it represents a correlation of encoded information rather than other more superficial signal features. Finally, the increasing use of noise signals in physiological experiments, either as direct neural stimulation or as auditory input, suggests that the correlative properties of noise-stimulated neurons are of interest to the neuroscience community.

A comprehensive overview of the properties and uses of pseudorandom sequences is provided in Golomb (1981), and Proakis (2000) explains their use in modern digital telecommunications.

In the system of Figure 9, we have used a 1023-bit code with bit length of 0.1 ms, so the code length is 0.1023s. The codes were shifted to have zero mean (i.e. bit levels of ±0.5) and amplitude scaled to a bit amplitude of 100μV. The code was repeated continuously.

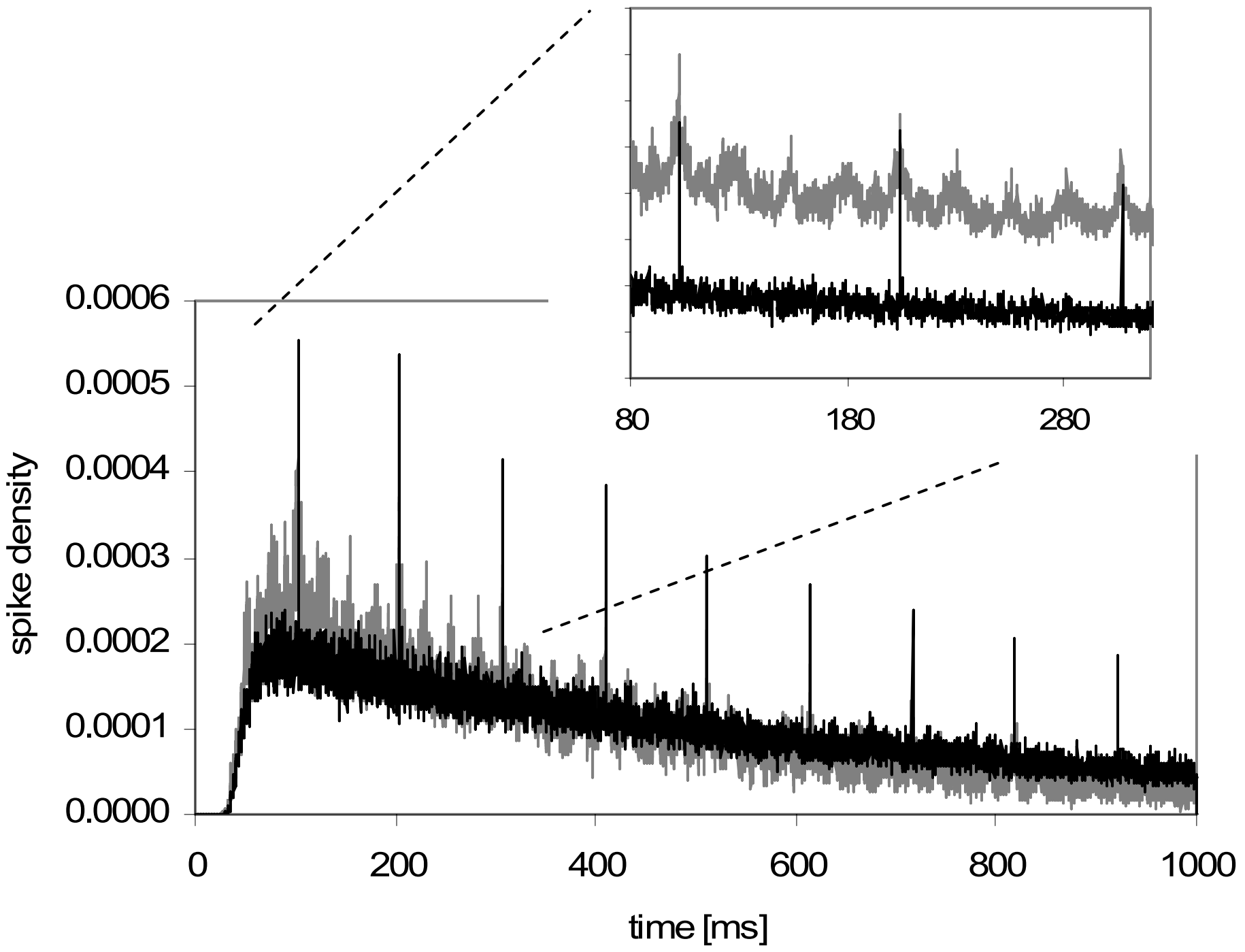

**Figure 9:** A comparison of the phase-continuous ISI obtained from the AM-ISI model (black) and simulated (gray) Ornstein-Uhlenbeck neuron is shown for a 1023-bit pseudorandom code input. It can be seen that the sharp peaks in the autocorrelation function are matched in the AM-ISI model and Ornstein-Uhlenbeck simulation; the inset shows the two curves displaced vertically for clarity. For a 1023-bit PRN code with bit period of 0.1 ms, we expect autocorrelation peaks at intervals of 102.3ms, as shown. Neuron parameters were $\tau_m$ = 10ms, $\theta$ = -54mV, $V_{reset}$ = -80mV, $V_L$ = -70mV, $\sigma$=0.1 mV/√Hz, $w$ = 3.00, with the signal scaled to a bit amplitude of 100μV. The mismatch in overall shape between the model and measured ISI is due to the noise-like nature of the PRN code, which effectively increases the noise parameter, thereby causing a higher, earlier peak in the ISI. The regular peaks in the measured ISI are an inaccuracy arising from the use of larger than appropriate signal amplitudes, to achieve visual clarity.

Figure 9 suggests that the presence of the autocorrelation function in the ISI is not an artifact of the signal envelope, or some other type of artifact; that the precision in time of the correlation is very high; and that the process outlined in (5)-(10) provides a reasonable description of the properties of the system dynamics.

## 3. Cross-Correlation Systems

One of the advantages of this model is the ease with which it allows single neuron response to small periodic signals to be predicted, both intuitively and mathematically. In this section, we will show how the model allows us to design and analyze two different neural systems for cross-correlating input signals.

We have seen in the above sections that single neurons can produce an output that contains the autocorrelation function of the input stimulus. Autocorrelation functions play a significant part in many neural processes; for example, psychoacoustic effects such as the pitch shift and missing fundamental effects can be explained if the auditory nerve is able to produce an autocorrelation function of the auditory input (Colburn, 1966). However, cross-correlations are much more widely useful in both biological neural processing and electrical engineering.

The standard method for cross-correlation in spiking neurons is the coincidence detector. In vision systems this is referred to as an enhanced motion detector (EMD) (Reichardt & Egelhaaf, 1988; Borst & Egelhaaf, 1989). EMDs are only capable of providing correlation output at fixed delays; the mechanisms which we will describe here give correlations over an extended range in time. Cross-correlation forms a basic process in many sensory systems – particularly the auditory and vision systems (Reichardt & Egelhaaf, 1988; Borst & Egelhaaf, 1989; Jeffares 1948; Colburn, 1996) – as well as in associative memory systems (Hassoun, 1993), so a robust cross-correlation circuit may have some utility in modeling these systems. The foundations of coincidence detection are set out by Srinivasan and Bernard (1976), in a paper which describes a mechanism of neural multiplication by a product of probabilities; it is this same mechanism, with the probabilities separated in time, that produces the correlations observed in the current work.

We show here two methods for producing cross-correlation functions using spiking neurons, which have emerged as a result of the explicit separation of autocorrelation and neuron dynamics offered by this model. These methods are inherently engineering circuits rather than biologically plausible neural systems, but we think they are important as a proof of physical feasibility. That is to say, they show that the autocorrelation and cross-correlation are not *ex post facto* artefacts of the mathematical analysis, but are inherent properties of the spike trains. We conclude by showing how the most useful circuit could be implemented as a physiologically-plausible four-neuron mutual inhibition network.

### 3.1 Cross-Correlation Extraction from the Autocorrelation Function

The requirement in this case is to correlate an unknown input signal with a known or reference signal. The resulting cross-correlation function gives an indication of the similarity between signals over a range of time delays. We can extract the cross-correlation function from the difference of the unknown and reference signals, say *f(t)* and *g(t)* respectively, as follows:

$$\begin{aligned} &R_{(f-g)(f-g)}(\tau) \\ &\quad = \frac{1}{T}\int_{-T/2}^{T/2}\left[(f(t)-g(t))\cdot(f(t+\tau)-g(t+\tau))\right]dt \\ &\quad = \frac{1}{T}\int_{-T/2}^{T/2}\left[(f(t)f(t+\tau)-f(t)g(t+\tau)-g(t)f(t+\tau)+g(t)g(t+\tau)\right]dt \\ &\quad = R_{ff}(\tau)-R_{fg}(\tau)-R_{gf}(\tau)+R_{gg}(\tau) \end{aligned} \tag{19}$$

For real-valued signals $R_{fg}(\tau) = R_{gf}(-\tau)$ so we do not get the cross-correlation out completely unambiguously; nonetheless there are many situations in which the following function is useful:

$$R_{ff}(\tau) + R_{gg}(\tau) - R_{(f-g)(f-g)}(\tau) = R_{fg}(\tau) + R_{fg}(-\tau). \tag{20}$$

As an example of how these signals can be extracted, we have constructed an electronic circuit (Tapson & Etienne-Cummings, 2007), consisting of two analog integrate-and-fire neurons and a digital accumulator, which is able to extract the function in (20) above. A block diagram of the circuit is shown in Figure 10. It consists of integrate-and-fire neurons constructed using simple operational amplifier integrators, with digital counters which count the time since the last firing event. The output of each neuron is indicated by an event line, which goes momentarily high when the neuron fires, and eight digital lines which carry the parallel output state of an 8-bit counter. These lines are fed into a differential histogram accumulator. Each time the upper neuron in Figure 10 fires, the accumulator bin corresponding to its counter output is incremented (for example, if the count is 233 when the neuron fires, bin 233 is incremented by one). Each time the lower neuron fires, the appropriate bin is decremented. The state of each bin of the differential accumulator therefore reflects the difference in interspike-interval densities between the two input signals, thereby directly implementing the subtraction as shown in Equation (21) which follows below.

We have tested this system extensively with a number of different signals, including sums of sinusoids, and digital signals such as the pseudorandom codes mentioned in Section 2.

The particular architecture used in this system is based on the expectation that one signal will be a reference signal, and the other an unknown signal, whose similarity and phase delay with respect to the reference is of interest. Actual electronic implementation is shown in Tapson & Etienne-Cummings (2007), and further results in Folowosele et al. (2007).

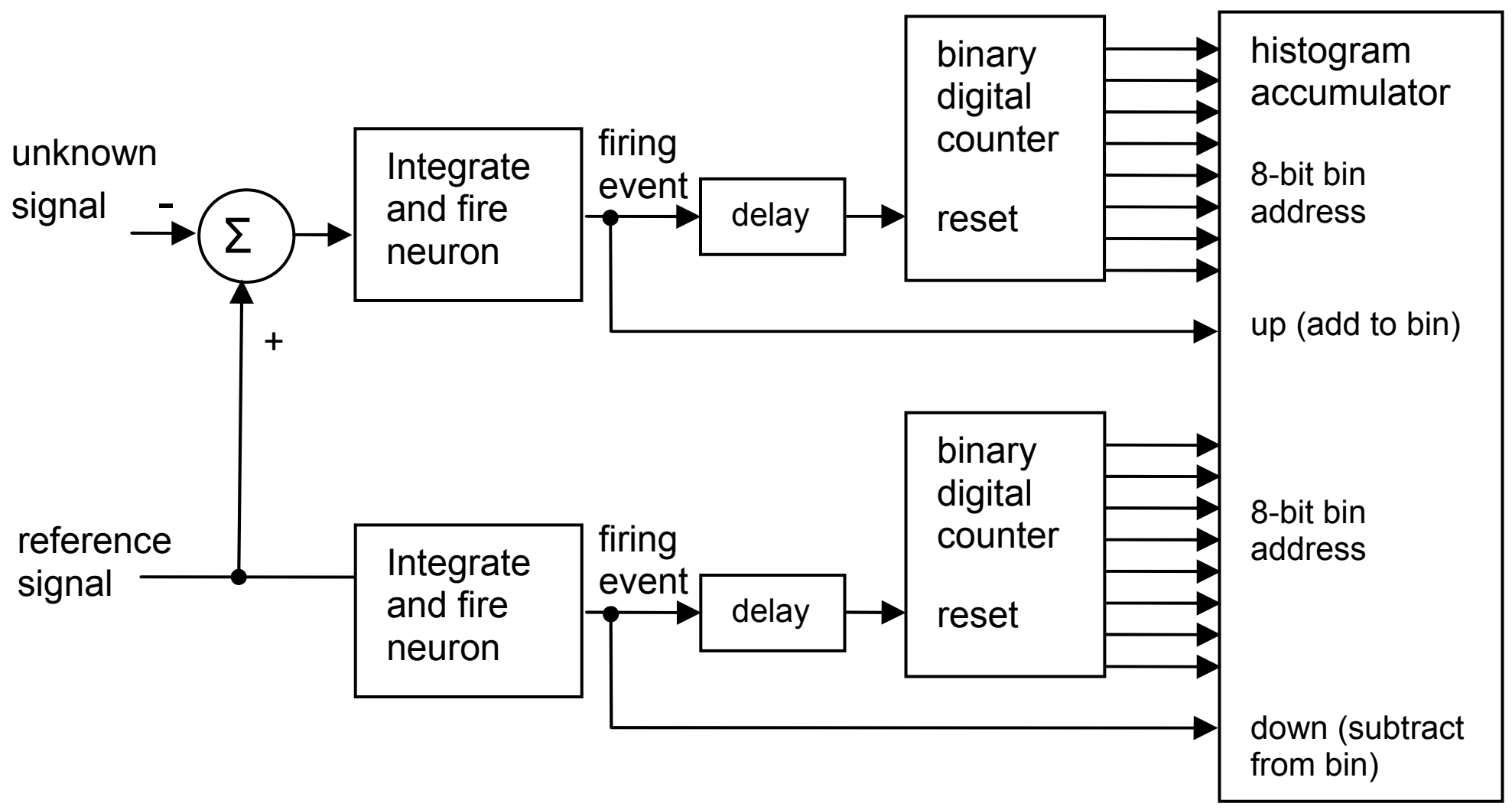

**Figure 10:** Architecture of a circuit which extracts cross-correlation terms for two signals using electronic spiking neurons and the approach in (19) - (21). For strictly correct implementation of (21) the "down" count should subtract two spikes for each firing event. Further details of an electronic implementation can be found in Tapson & Etienne-Cummings (2007).

In Figure 11 we demonstrate how this arrangement would extract the cross-correlation products from a smooth, more real-world signal. The signal is shown in the top panel of Figure 11; it is a sum of five sine waves, with frequencies taken randomly from a harmonic series of ten possible frequencies, with random amplitudes and phases. This signal was cross-correlated with a copy of itself, delayed by a fixed time (150 ms in this example). We call the signal $g(t)$ and the delayed copy $g(t+\theta)$ in the analysis which follows.

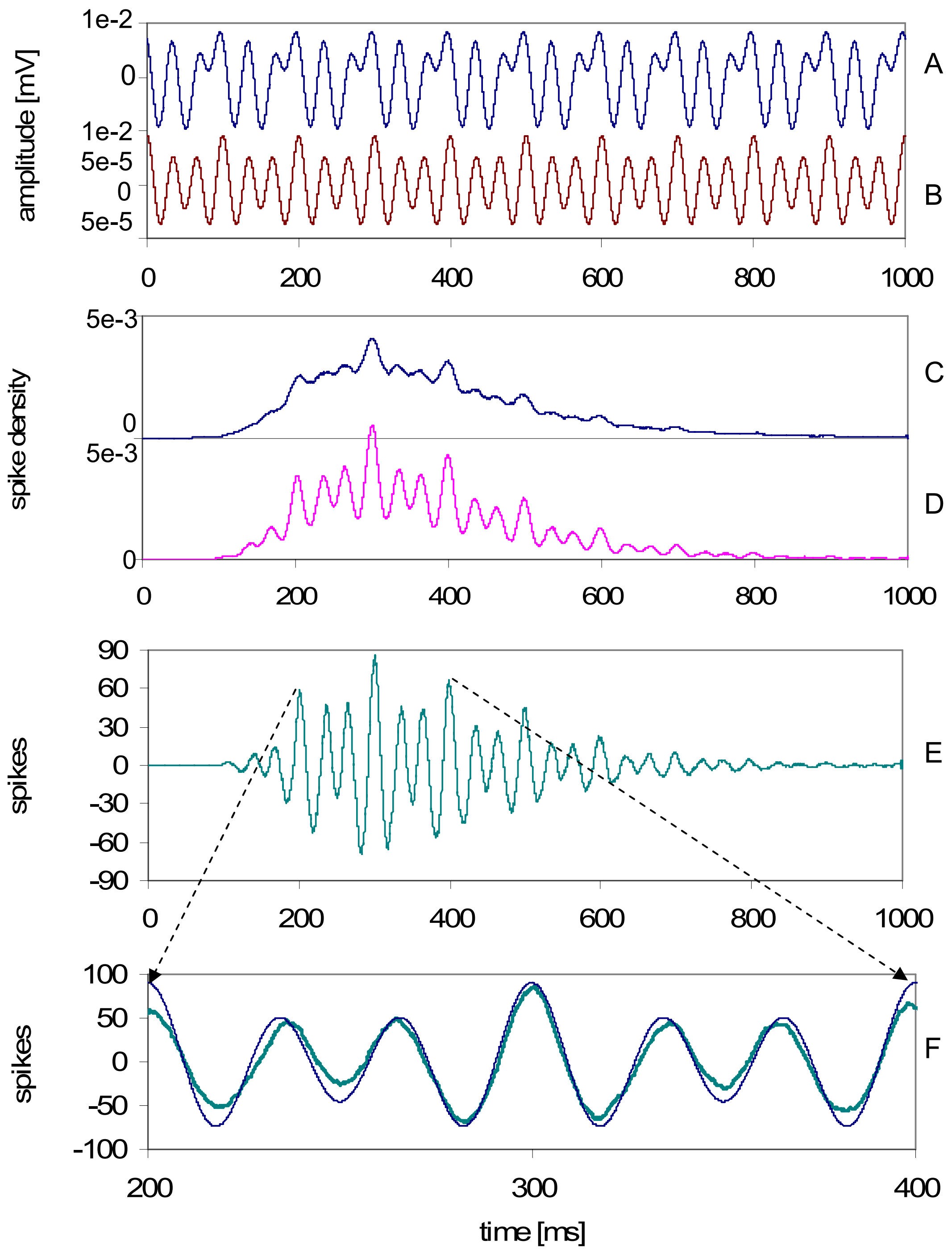

**Figure 11:** Signals in the extraction of cross-correlation components by a system such as that in Fig 10. Curve A shows the signal $g(t)$ (constructed as described in the text). Curve B shows the cross-correlation components $R_{g(t)g(t+\theta)}+R_{g(t+\theta)g(t)}$ calculated mathematically. Curves C and D show the ISIs of the neurons receiving the reference

*g*(*t*) (D) and (reference minus unknown) *g(t)-g(t+θ)* signals (C). Curve E shows the differential ISI. Curve F shows the center portion of the differential ISI (bold line), with the cross-correlation curve B (fine line) scaled and overlaid for comparison. The horizontal scale is time in ms. The neurons were simple I&F models with $\sigma$=0.1 mV/√Hz, $\theta$=15 mV, $m$= 40 mV/s.

The model is used to predict the output of the system shown in Figure 10 as follows: the outputs of the neurons are

$$
\begin{aligned}
\rho_{1,AM}(\tau) &= \rho(\tau)[1+2R_{g(t)g(t)}(\tau)] \\
\rho_{2,AM}(\tau) &= \rho(\tau)\left[1+R_{[g(t)+g(t+\theta)][g(t)+g(t+\theta)]}(\tau)\right] \\
\rho_{2,AM}(\tau)-\rho_{1,AM}(\tau) &= \rho(\tau)\left[1+R_{[g(t)+g(t+\theta)][g(t)+g(t+\theta)]}(\tau)-1-2R_{g(t)g(t)}(\tau)\right] \qquad (21) \\
&= \rho(\tau)\left[R_{[g(t)+g(t+\theta)][g(t)+g(t+\theta)]}(\tau)-2R_{g(t)g(t)}(\tau)\right] \\
&= \rho(\tau)\left[R_{g(t)g(t+\theta)}(\tau)+R_{g(t+\theta)g(t)}(\tau)\right]
\end{aligned}
$$

This final result may be applied to predict the differential ISI curve in the lowest two panels of Figure 11. It can be seen that there is a good match for practical purposes between the model and the simulated results, and that this system can be used to extract the cross-correlation terms from the two input signals.

### 3.2 Direct Cross-Correlation

Consider the (somewhat contrived) arrangement of two integrate-and-fire neurons in Fig 12. This architecture has been designed so that each neuron remains in an inhibited state after firing, until the other neuron fires.

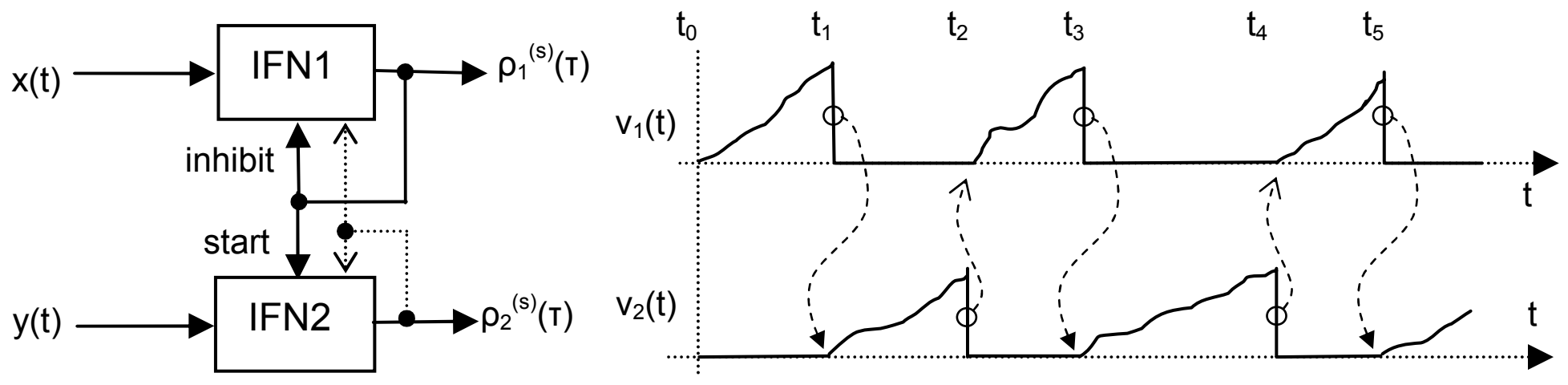

**Figure 12:** A simple network of two neurons, in which neuron IFN1 is stimulated by signal $x(t)$, and neuron IFN2 by signal $y(t)$. Each spike from IFN1 starts the next cycle of IFN2, and inhibits IFN1 until IFN2 spikes. Each spike from IFN2 starts a cycle of IFN1, and inhibits IFN2 until IFN1 spikes. The potentials on the right show the sequence of integrate-and-fire events (with inhibit/start dependencies marked with dotted arrows).

We can use our AM-ISI model to predict the firing distributions of the neurons, as follows. We consider the probability densities for the sequence of firing where IFN2 fires at $t_0$ (and is thereafter inhibited); IFN1 fires at $t_1$, disinhibiting IFN2; and then IFN2 fires at $t_1 + \tau$. Note that the neurons are constrained to spike alternately by the architecture. The neurons have the same unstimulated interspike-interval distribution $\rho(\tau)$, but their conditional interval distributions $\rho_{n1}(\tau|t)$ and $\rho_{n2}(\tau|t)$ are different because they are subject to different stimuli.

$q_{2(IFN2),1(IFN1)}(\tau, t_{0(}\text{(IFN1)}) = \{\text{Prob that in}$

$\text{sequence}\,[t_0\text{(IFN2)}, t_1\text{(IFN1)}, t_1\ \text{(IFN2)}], t_1 \text{ and } t_2 \text{ are separated by } \tau\}$

$$= \int_{t_0}^{\infty} \rho_{IFN1}(t_1 + \tau | t_1) \rho_{IFN2}(t_1 | t_0) dt_1$$

$$= \int_{t_0}^{\infty} \rho(\tau)\rho(t_1 - t_0)(1 + wy(t_1 + \tau))(1 + wx(t_1))dt_1$$

$$= \int_{t_0}^{\infty} \rho(\tau)\rho(t_1 - t_0)dt_1 + w\int_{t_0}^{\infty} \rho(\tau)\rho(t_1 - t_0)y(t_1 + \tau)dt_1$$

$$+ w\int_{t_0}^{\infty} \rho(\tau)\rho(t_1 - t_0)x(t_1)dt_1 + w^2 \int_{t_0}^{\infty} \rho(\tau)\rho(t_1 - t_0)y(t_1 + \tau)x(t_1)dt_1 \quad (22)$$

$$\approx \rho(\tau) + w^2 \rho(\tau)\int_{t_0}^{\infty} y(t_1 + \tau)x(t_1)dt_1$$

$$= \rho(\tau)\left(1 + w^2 R_{xy}(\tau)\right) \quad (23)$$

The cross-correlation function for signals $x(t)$ and $y(t)$ is present explicitly in (23). Note that this interval distribution on $\tau$ is not a standard interspike interval, but represents a joint interspike-interval distribution of IFN1 and IFN2. The cross-correlation can be extracted by a differential histogram, similar to that shown in the circuit in Figure 10.

In order to demonstrate this circuit, we have used it in a typical CDMA situation, where the amplitude and phase of a particular Gold code has to be extracted from a signal containing several codes. In this case, *x(t)* would be our input signal, which is a linear sum of several Gold codes; and *y(t)* would be the reference signal, containing only one code. The cross-correlation function should contain only the cross-correlation peak indicating the amplitude and phase of the matching Gold code in *x(t)*.

Figure 13 shows the autocorrelation function of *x(t)*, which is a linear mixture of two Gold codes; this would represent a received signal consisting of a code of interest, delayed by some unknown time or phase $\Delta\theta$, mixed with an unwanted, interfering code. The cross-correlation function of *x(t)* and *y(t)* is shown; this would represent the desired product of a demodulation, as it would indicate the amplitude and phase of the code in *x(t)* which we desire to detect; and the outputs of both of the neurons. It can be seen that the neurons have extracted the desired amplitude and phase, with opposite polarities of phase delay $\Delta\theta$ as expected. This shows that the system is able to unambiguously extract the cross-correlation components from the two input signals.

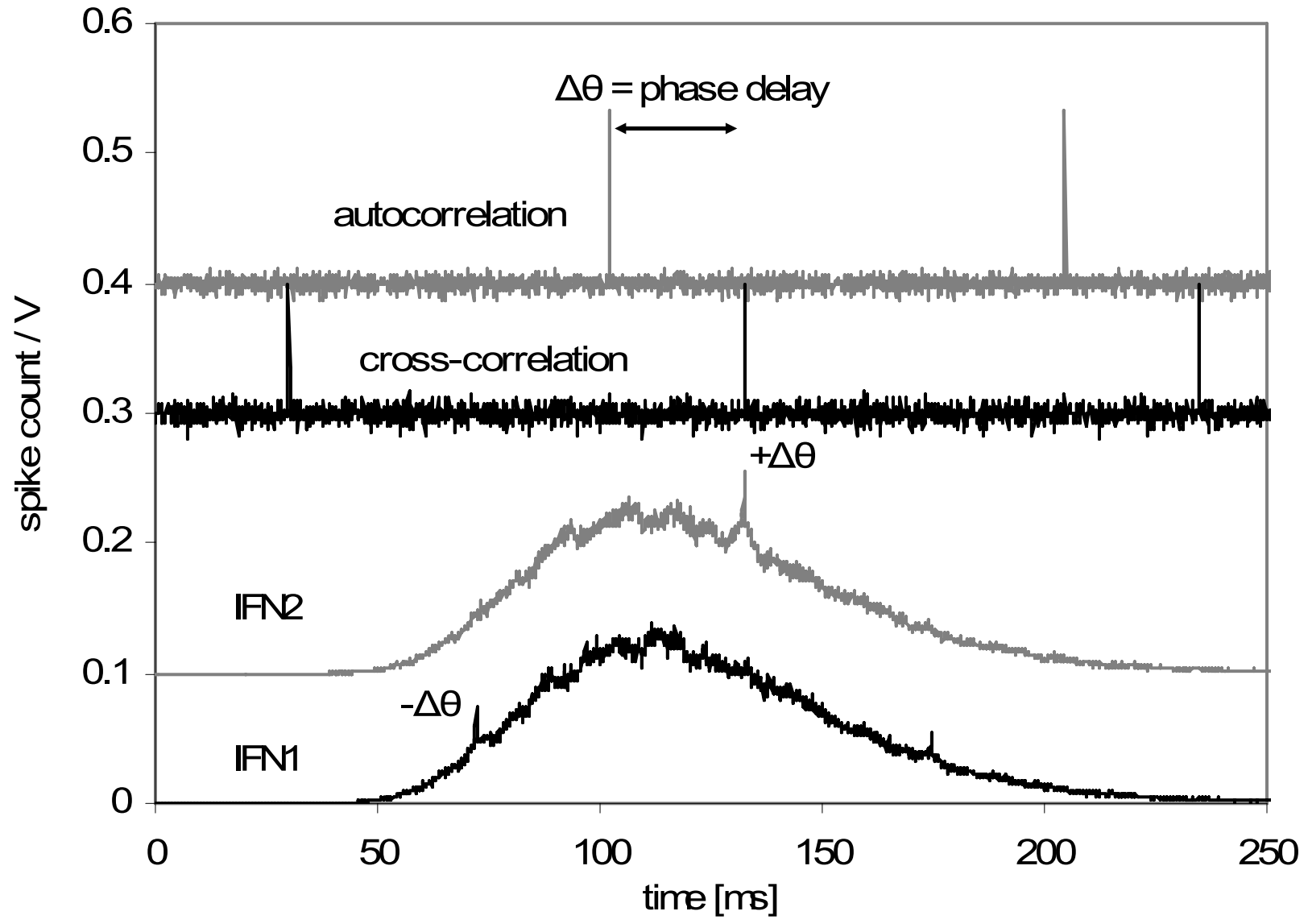

**Figure 13:** This shows, from top to bottom: autocorrelation function of the input signals *x(t)* and *y(t)* (they have identical autocorrelation functions); cross-correlation function of *x(t)* and *y(t)*; the stationary interspike-interval density of IFN2; the stationary interspike-interval density of IFN1. It can be seen the IFN outputs do not include the autocorrelation function, but only (and separately) the positive and negative cross-correlation terms for the Gold code that was common to both the input and reference signals. Curves have been displaced vertically for clarity. Note that because the model output of IFN1 includes $R_{xy}(\tau)$, and the output of IFN2 includes

$R_{yx}(\tau)$ = - $R_{xy}(\tau)$, the sign of the phase delay is opposite in their outputs. The phase delay was 300 bits in a 1023 bit code. Neurons were simple I&F models with $\sigma$=1.25 mV/√Hz, $\theta$=1000mV, $m$= 7.5 mV/ms.

### 3.3 Physiological Implementation

It might be considered that the circuits shown above are not physiologically plausible. We can recast the circuit of Figure 12 in the form of two reciprocal inhibition networks combined into a four-neuron mutual inhibition network (Matsuoka, 1987) using physiologically plausible neurons, as shown in Figure 14. Combinations of half-center oscillator networks such as these are believed to form central pattern generators (CPGs) in animal locomotion, for example (Cohen et al., 1988; Grillner et al., 1998; Grillner, 2003).

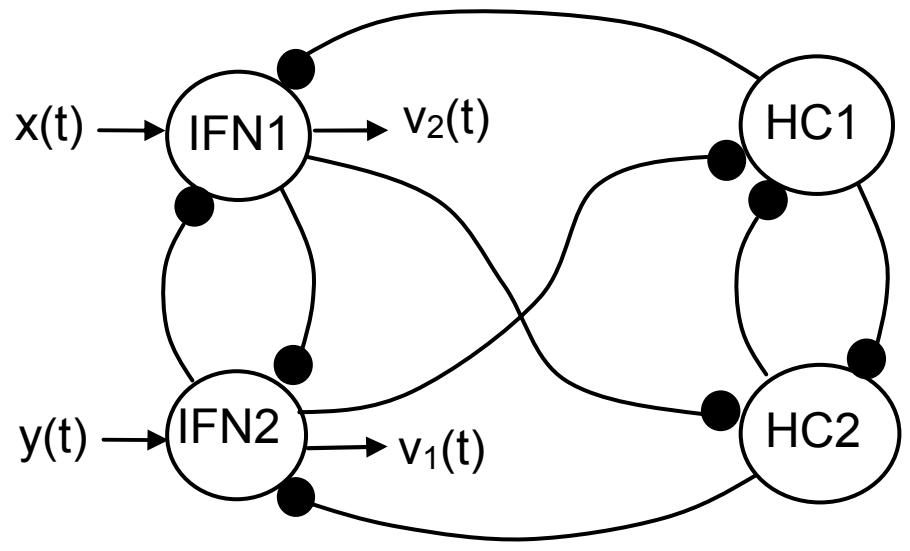

**Figure 14.** A four-neuron mutual inhibition network constructed from two reciprocal inhibition networks (half center oscillators). All connections are inhibitory. HC1 and HC2 are implemented to give bursting behavior, so that they provide sustained inhibition to IFN1 and IFN2 alternately. If HC1 is bursting, HC2 and IFN1 are inhibited. When IFN2 integrates and fires, this halts the bursting of HC1, allowing HC2 to begin bursting and thereby sustaining the inhibition of HC1 and IFN2 until

IFN1 fires. The signals $x(t)$ and $y(t)$, and $v_1(t)$ and $v_2(t)$ correspond to those in Figure 12.

The circuit of Figure 14 requires little fine-tuning to work; the most important tuning issue is that the neurons IFN1 and IFN2 should have a hyperpolarization limit, so that when they are disinhibited at the start of an integrating cycle, the membrane potential starts from a consistent level. The purpose of the inhibitory connections between IFN1 and IFN2 is to ensure that each is at this level immediately following a firing event of the other, thereby ensuring that the integration starts at a time and level determined by the firing event of the other.

Figure 15 shows typical waveforms from a network of the type shown in Figure 14.

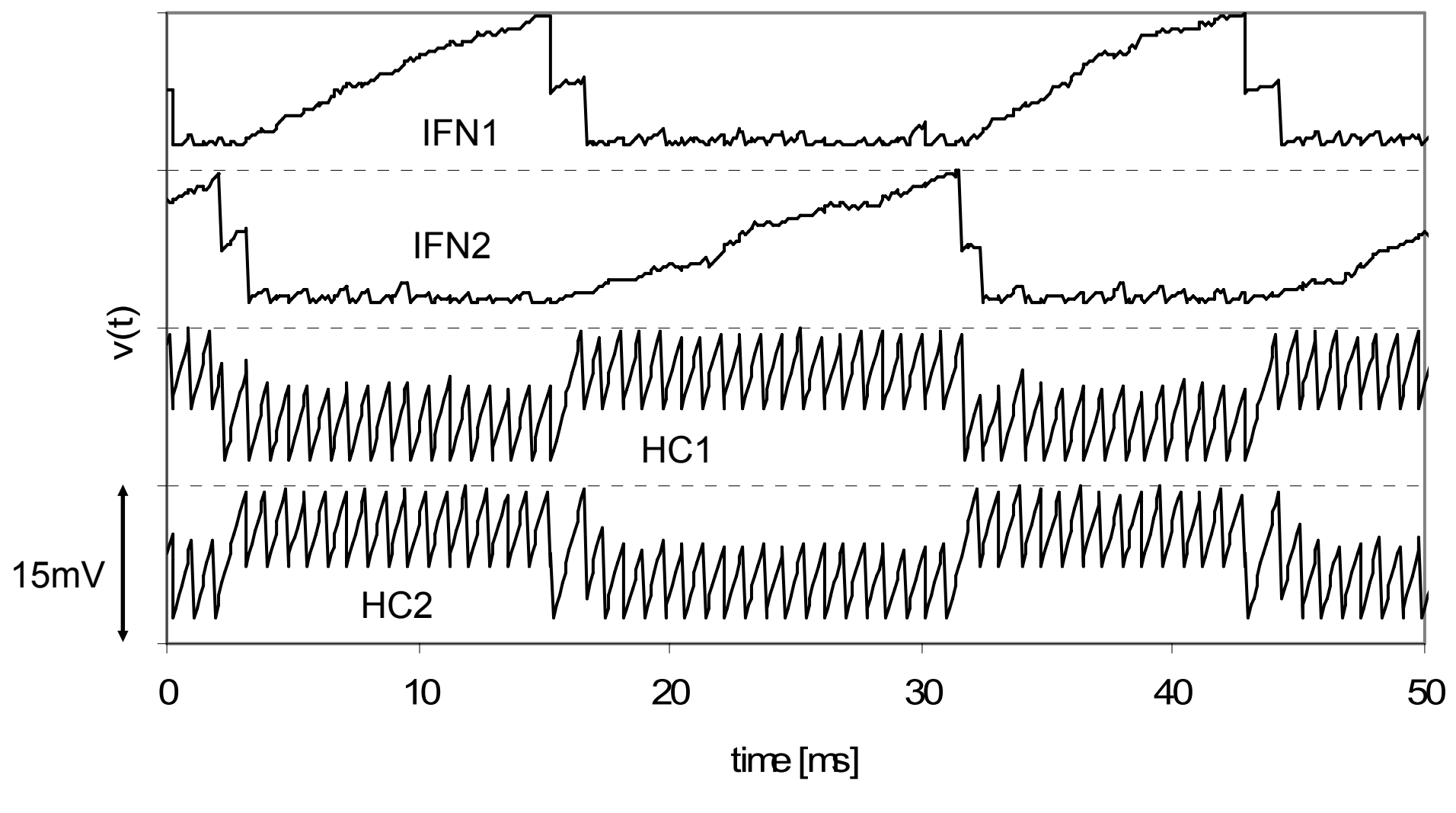

**Figure 15:** Typical trajectories of the membrane potentials $v(t)$ for the neurons IFN1, IFN2, HC1 and HC2 as shown in Figure 14. The curves have been displaced vertically for clarity. The vertical scale is indicated by the marker at left, and the threshold value for each neuron is indicated by a dotted line. Each neuron had a hyperpolarizing barrier $V_{hyper} = \theta$-12.5 mV. It can be seen that neuron HC1 maintains

IFN1 in a hyperpolarized state until IFN2 fires; the additional inhibitory connection from IFN2 to IFN1, shown in Figure 14, ensures that IFN1 begins integrating at a consistent time and potential relative to the firing of IFN2. Similarly, IFN2 is inhibited by HC2, with additional inhibition provided by the firing of IFN1. The neurons were based on the standard OU model, with network equations and parameters given in Appendix I.

The output from a circuit implemented as in Figures 14 and 15 is indistinguishable from that in Figures 12 and 13, provided that care is taken with the implementation. Note however that the firing intervals are no longer first-order intervals for a single neuron, but first-order intervals for alternate spikes of IFN1 and IFN2 (this applies to both the Figure 12 and Figure 14 implementations).

The circuits and results shown in Figures 10-15 demonstrate that we are able to construct useful neural circuits using an intuitive approach which is enabled by the simplicity of the model described in this paper. In addition, it throws some light on a possible method for cross-correlation in physiological neural systems.

### 3.3.1 Downstream Processing of Interspike Interval Outputs

While the differential histogram circuit shown in Figure 10 provides a means for useful extraction of the cross-correlation in an engineering application, it remains to be confirmed whether ISI-based representation can be found in the nervous system. For example, it is known that cross-correlation is represented in the auditory system (in the inferior colliculus) for sound localization (Hancock & Delgutte, 2004;

Shackleton et al., 2005). Furthermore, there is evidence that cross-correlation also plays a role in pitch detection (Cedolin & Delgutte, 2005). For the latter, it is postulated that a functional ISI must be computed by populations of neurons in order to have a place code for pitch, where individual neurons in the population are broadly tuned to a preferred ISI and the response of said neurons represents the frequency of occurrence of the preferred ISI. Moreover, Arbarbanel & Salathi (2006) have recently demonstrated a neural circuit based on Hodgkin-Huxley neurons that is capable of detecting specific ISI sequences. Whether these biological systems can perform the specific ISI computation proposed in this paper is still an open question, however, for detection of simple peaks of significant salience in the ISI, the Arbarbanel & Salathi method should suffice.

## 4. Conclusions

We have presented a simple approximate model for the interspike-interval density of a continuously stimulated integrate-and-fire neuron, which acknowledges the Markov nature of the response. The model allows both the conditional interspike-interval density and the phase-continuous stationary interspike-interval density to be expressed as products of a term involving only the neural characteristics, and a term involving only the signal characteristics. The approximation shows particularly clearly that signal autocorrelations and cross-correlations arise as natural features of the interspike-interval density, particularly for small signals and moderate noise, in a wide range of neural types.

## Appendix I: Simulation and Implementations

### A 1. General Simulation Methods

Neural models were simulated using MATLAB. Given the wide number of neuron types modelled, and the importance of noise in the model, it was decided to avoid using variable-timestep algorithms for integration, but rather to use a simple rectangular approximation $x(t+\Delta t) = x(t) + \Delta t(dx/dt)$ with sufficiently small timesteps to ensure accuracy and stability (and in particular, to ensure sufficient noise bandwidth). Timesteps of 1 and 10 μs were used according to the situation, with consistent use of one timestep for each model. The typical update expression for an OU neuron membrane, corresponding to (17) would therefore be of the form:

```
v=v-(v-vl)/tm+randn*sigma+signal(mod(t,T)+1);
```

where `v` is the membrane potential, `vl` the reversal potential, `tm` the membrane time constant (in terms of the timestep, not real time), `randn` a MATLAB function which generates normally-distributed random numbers with unity variance, `sigma` the noise amplitude, `signal` the signal vector, `t` the time index, and `T` the length of the signal vector (usually the period of the signal fundamental). The noise and signal would be in voltage form, hence there being no need to scale them with membrane impedances. The signal would be prescaled prior to the operation above, and have any DC drift component added at the same time, to reduce the number of iterated calculations.

When a baseline ISI for the unstimulated neuron was required, these were calculated using (16) for I&F neurons, and generated by simulation in the case of OU and HH neurons. A series of tests were performed to compare curves from (16) with simulated unstimulated I&F neurons, in order to provide a measure of the number of spikes required in a simulation to provide a reasonably smooth and accurate baseline ISI. It was found that ISIs of $10^7$ spikes would reliably give errors of $E < 10^{-3}$, as would ISIs of $10^6$ spikes if these were post-filtered with a simple sliding window average across bins. This conclusion was tested for OU and HH neurons by measuring the error between baseline ISIs for identical neurons, e.g. by measuring the errors between two supposedly identical ISIs. In calculating errors using (16), it was assumed that a baseline ISI with an error $E < 10^{-3}$ was sufficiently accurate for calculating model errors of $E \geq 10^{-2}$, i.e. an order of magnitude margin of accuracy between the baseline and the tested ISI was considered to be adequate.

When measuring ISIs for stimulated neurons, in most cases the simulation was run until approximately $10^6$ spikes had accumulated in the ISI, although a numerical test for "smoothness" of the ISI, based on rms variation between adjacent bins, was also used in some cases to terminate simulation. For the data used for figures in this paper, visual smoothness was often used as a criterion for terminating the simulation.

**A 2. The Hodgkin-Huxley Model**

The Hodgkin-Huxley model used in this work was that described in Dayan & Abbott, 2001, as follows:

$$c_m \frac{dv(t)}{dt} = -i_m + \frac{I_e}{A}$$

where $c_m$ is the membrane capacitance per unit area, $i_m$ is the membrane current per unit area, $I_e$ is the electrode current (which we can use to denote a generic external current source or sink) and $A$ is the membrane area. The membrane current depends on the leakage and ionic currents, conductances ($g$) and reversal potentials ($E$):

$$i_m = \bar{g}_L(v - E_L) + \bar{g}_K n^4(v - E_K) + \bar{g}_{Na} m^3 h(v - E_{Na})$$

where the subscripts *L*, *K* and *Na* refer to leakage, potassium and sodium channels respectively. *n, m* and *h* are activation variables governed by:

$$\frac{dx}{dt} = \alpha_x(v)(1 - x) - \beta_x(v)x$$

where $x \in (m,n,h)$ is the relevant activation variable. The specific rate functions *α*, *β* are:

$$\alpha_n(v) = \frac{0.01(v + 55)}{1 - \exp(-0.1(v + 55))}$$

$$\beta_n(v) = 0.125\exp(-0.0125(v + 65))$$

$$\alpha_m(v) = \frac{0.1(v + 40)}{1 - \exp(-0.1(v + 40))}$$

$$\beta_m(v) = 4\exp(-0.0556(v + 65))$$

$$\alpha_h(v) = 0.07\exp(-0.05(v + 65))$$

$$\beta_n(v) = \frac{1}{1 + \exp(-0.1(v + 35))}$$

for $\alpha_x$, $\beta_x$ in $ms^{-1}$ and *v* in mV.

The conductance, reversal potential and membrane parameters used were $g_L$ = 0.003 mS/ $mm^2$, $g_K$ = 0.036 mS/ $mm^2$, $g_{Na}$ = 1.2 mS/$mm^2$, $E_L$ = -54.3mV, $E_K$ = -77mV, $E_{Na}$ = 50mV, $A$ =$1.26\times10^{-5}$ $cm^2$ and $c_m$ = 1 μF/$cm^2$. The initial values of *v*, *n*, *m* and *h* were -65mV, 0.318, 0.053, and 0.595 respectively.

In simulation, the differential equations were recast as difference equations with a timestep of 10μs. For both the conditional and phase-continuous simulation the neuron was never reset after spiking, but simply allowed to free-run after the initial start. A threshold of -15mV was used to define the spiking level; in conditional simulation, the signal (but not the neuron) was reset to its starting phase after an upward transition across this threshold. For continuous stimulation, a spike event was recorded at the time of transition, but neither signal nor neuron were reset at any stage.

### A 3. The Four-Neuron Mutual Inhibition Network

The four-neuron mutual inhibition network was modeled using the framework provided by Matsuoka (1987), using OU neurons. The network is described using vectors of weights and a connection matrix as follows:

$$\boldsymbol{\tau}_{\mathbf{m}} \frac{d\mathbf{v}}{dt} = -\mathbf{v} + \mathbf{s} - \mathbf{a} \times \mathbf{y} + \boldsymbol{\sigma}\boldsymbol{\zeta} + \mathbf{A}\mathbf{g} + \mathbf{m}$$

where $s$ is the vector of reversal potentials (scaled for impedance), **a** is the neuron connection matrix, **y** is the neuron output (spike) vector, and all the other variables have their usual meaning, but are vectors rather than scalars. The values used for the network were:

$$\boldsymbol{\tau}_{\mathbf{m}} = \begin{bmatrix} 100 \\ 100 \\ 10 \\ 10 \end{bmatrix} \text{ms}, \quad \mathbf{s} = \begin{bmatrix} 2 \\ 2 \\ 1 \\ 1 \end{bmatrix} \text{mV}, \quad \mathbf{a} = \begin{bmatrix} 0 & 2 & 2 & 0 \\ 2 & 0 & 0 & 2 \\ 0 & 2 & 0 & .11 \\ 2 & 0 & .11 & 0 \end{bmatrix}, \quad \boldsymbol{\sigma} = \begin{bmatrix} .002 \\ .002 \\ .001 \\ .001 \end{bmatrix} \text{mV}/\sqrt{\text{Hz}},$$

$\mathbf{m} = \begin{bmatrix} 10 \\ 10 \\ 100 \\ 100 \end{bmatrix}$ mV/ms, $\mathbf{A} = \begin{bmatrix} .25 \\ .25 \\ .25 \\ .25 \end{bmatrix}$ mV. The output spike matrix **y** had elements of unity if a spike occurred for the corresponding neuron in a given timestep, or zero otherwise.

This network was used successfully, but incurred occasional errors as a result of having two neurons spike simultaneously in the same timestep. The precedence of apparently simultaneous spikes was forced by adjusting **a** as follows:

$$\mathbf{a} = \begin{bmatrix} 0 & 2 & 2 & 0 \\ 2 & 0 & 0 & 2 \\ -0.05 & 2 & 0 & .11 \\ 2 & -0.05 & .11 & 0 \end{bmatrix}.$$

This violates the inhibition-only structure of the network; the alternative would be to increase the timestep precision or tolerate the small number of errors.